\documentclass[journal]{IEEEtran}
\usepackage{cite}
\usepackage{siunitx}
\ifx\pdfoutput\undefined
\usepackage{graphicx}
\else
\usepackage[pdftex]{graphicx}

\ifCLASSINFOpdf
\else
\fi
\usepackage{amsmath}
\usepackage{url}
\usepackage{multirow}

\begin{document}
%\title{Static and Dynamic Pull-in Behavior   of  Hybrid Levitation Micro-Actuators: Simulation, Modelling and Experimental Study}
\title{Static  Pull-in Behavior   of  Hybrid Levitation Micro-Actuators: Simulation, Modelling and Experimental Study}

%\author{Michael~Shell,~\IEEEmembership{Member,~IEEE,}
%        John~Doe,~\IEEEmembership{Fellow,~OSA,}
%        and~Jane~Doe,~\IEEEmembership{Life~Fellow,~IEEE}% <-this % stops a space
%\thanks{M. Shell was with the Department
%of Electrical and Computer Engineering, Georgia Institute of Technology, Atlanta,

\author{Kirill~V.~Poletkin%
       \thanks{K. Poletkin is with the Institute of Microstructure Technology, Karlsruhe Institute of
Technology, Hermann-von-Helmholtz-Platz 1, 76344 Eggenstein-Leopoldshafen, Germany
%and with the Institute of Robotics and Computer Vision,  Innopolis University, 1 Universitetskaya street, Innopolis City, 420500, Russian Federation
e-mail: kirill.poletkin@kit.edu and kirill.v.poletkin@gmail.com.}}


\markboth{}%
{Poletkin \MakeLowercase{\textit{et al.}}}




\maketitle

\begin{abstract}
In this article, a systematic and comprehensive approach based on finite element analysis and analytical modelling for studying static   pull-in phenomena in hybrid levitation  micro-actuators is presented. A finite element model of electromagnetic levitation micro-actuators based on the Lagrangian formalism is formulated and developed as a result of recent progress in the analytical calculation of mutual inductance between  filament loops. In particular, the developed  finite element model allows us to calculate accurately and efficiently a distribution of  induced eddy current within a levitated micro-object. At the same time, this fact provides a reason for formulating the analytical model in which the distribution of the induced eddy current can be approximated by one circuit represented by a circular filament.
In turn, both  developed  models predict the static  pull-in parameters of hybrid levitation  micro-actuators  without needs for solving nonlinear differential equations. The results of modelling obtained by means of the
developed quasi-finite element and analytical model are verified by the comparison with experimental results.
\end{abstract}


% \fi
\IEEEpeerreviewmaketitle

\section*{Nomenclature}
\addcontentsline{toc}{section}{Nomenclature}
\begin{IEEEdescription}
%\item[$J_{x}$, $J_{y}$, $J_{z}$ ] Rotating coordinate frame.
[\IEEEsetlabelwidth{DLMA }]
\item[${\boldsymbol{\underline{e}}^X}$] vector base of fixed frame \{$X_k$\}
\item[${\boldsymbol{{e}}^X_k}$] $k$-unit vector of fixed frame \{$X_k$\} ($k=1,2,3$)
\item[${\boldsymbol{\underline{e}}^x}$] vector base of coordinate frame \{$x_k$\}
\item[${\boldsymbol{^{(s)}\underline{e}}^y}$] vector base of coordinate frame \{$^{(s)}y_k$\} $(s=1,\ldots,n)$
\item[${\boldsymbol{^{(j)}\underline{e}}^z}$] vector base of coordinate frame \{$^{(j)}z_k$\} $(j=1,\ldots,N)$
\item[$F_l$] generalized force ($l=1,2,3$) (\si{\newton})
\item[${\boldsymbol{g}}$]  gravity acceleration vector (\si{\meter\per\second^2})
\item[$h_l$] height of levitation (\si{\meter})
\item[$h$] space between the electrode surface and cm of levitated disc (\si{\meter})
\item[$i_{cj}$]  current in the $j$-wire loop $(j=1,\ldots,N)$, (\si{\ampere})
\item[$i_{s}$]  current in the $s$-circular element $(s=1,\ldots,n)$ (\si{\ampere})
\item[$K$] kinetic energy (\si{\joule})
\item[$N$] number of wire loops
\item[$n$] number of finite elements
\item[$L$] Lagrange function (\si{\joule})
\item[$L_{jj}^{c}$] self-inductance of the $j$-wire loop (\si{\henry})
\item[$L^{o}$]  self-inductance of the finite circular  element (\si{\henry})
\item[$L_{ks}^{c}$] mutual inductance between $k$-  and $s$-finite circular elements  (\si{\henry})
\item[$M_{kj}$]  mutual inductance between  the $k$-circular element and  the $j$-wire loop  (\si{\henry})
\item[$m$] mass of levitated object (\si{\kilogram})
%\item[\smash{\begin{IEEEeqnarraybox*}[][t]{l}
%J_{x},J_{y},J_{y}\\
%%\hphantom{J_{x},{}}J_{z}
%\end{IEEEeqnarraybox*}}] Central principal
%moments of inertia of rotor about the $x$, $y$, $z$ axes, kg$\cdot$m$^2$.
%\mbox{}
%[\IEEEsetlabelwidth{$XYZ$}]
%\item[\smash{\begin{IEEEeqnarraybox*}[][t]{l}
%XYZ\\
%%\hphantom{V_1,{}}V_3
%\end{IEEEeqnarraybox*}}] Fixed coordinate frame related to the gyroscope case.
%\mbox{}
\item[$\boldsymbol{q}$] translational position vector of levitated object (\si{\meter})
\item[$R$] electrical resistance of finite element (\si{\ohm})
\item[$R_e$] radius of finite element (\si{\meter})
%\item[$R_s$] radius of stabilization coil (\si{\meter})
\item[$R_{cj}$] radius of j-circular coil filament (\si{\meter})
\item[$^{(j)}\boldsymbol{r}_c$] linear position vector of $j$-wire loop (\si{\meter})
\item[$\boldsymbol{r}_{cm}$] linear position vector of centre of mass of levitated object (\si{\meter})
\item[$T_l$] generalized torque ($l=1,2,3$) (\si{\newton\meter})
\item[$W_m$] energy stored within electromagnetic field (\si{\joule})
\item[\{$X_k$\}] fixed frame  ($k=1,2,3$)
\item[\{$x_k$\}] coordinate frame attached to levitated object ($k=1,2,3$)
\item[\{$^{(s)}y_k$\}] coordinate frame attached to $s$-finite element ($k=1,2,3$)
\item[\{$^{(j)}z_k$\}] coordinate frame attached to $j$-wire loop ($k=1,2,3$)
\item
\item[\textit{Greek symbols}]
\item[$\beta$] dimensionless square voltage
\item[$\kappa$] dimensionless parameter $h/h_l$
\item[$\lambda$] dimensionless displacement $q_3/h$
\item[$\mu_l$] damping coefficient ($l=1,2,3$) (\si{\newton\second\per\meter})
\item[$\nu_l$] damping coefficient ($l=1,2,3$) (\si{\newton\second\per\radian})
\item[$\xi$] dimensionless parameter $h_l/(2R_l)$
%\item[$\Pi$] Potential energy, J.
\item[$\Pi$] potential energy (\si{\joule})
\item[$^{(s)}\boldsymbol{\rho}$] vector of linear position of $s$-circular element in vector base ${\boldsymbol{\underline{e}}^x}$ (\si{\meter})
\item[$\Psi$] dissipation energy (\si{\joule\per\second})
\item[$^{(s)}\boldsymbol{\phi}$] vector of angular position of $s$-circular element (Brayn angles) (\si{\radian})
\item[$^{(j)}\boldsymbol{\phi}_c$] vector of angular position of $j$-wire loop (Brayn angles) (\si{\radian})
 \item[$\boldsymbol{\varphi}$] angular position vector of levitated object (\si{\radian})
 \item[$\boldsymbol{\omega}$] vector of angular velocity of levitated object (\si{\radian\per\second})
 \item
 \item[\textit{Subscripts}]
 \item[cm] centre of mass
 \item[DLMA] diamagnetic levitation micro-actuators
\item[ELMA] electric levitation micro-actuators
\item[FEM] finite element model
\item[HLMA] hybrid levitation micro-actuators
\item[ILMA] inductive levitation micro-actuators
\item[MLMA] magnetic levitation micro-actuators
 \end{IEEEdescription}

\section{Introduction}
%\IEEEPARstart{T}{his} demo file is intended to serve as a ``starter file''
%for IEEE journal papers produced under \LaTeX\ using
%IEEEtran.cls version 1.8b and later.
%% You must have at least 2 lines in the paragraph with the drop letter
%% (should never be an issue)
%I wish you the best of success.

\IEEEPARstart{E}{lectromagnetic} levitation micro-actuators employing remote ponderomotive forces,  in order to act on a target environment or simply compensate a gravity force for holding stably a micro-object at the equilibrium without mechanical attachment, have   already found wide applications and demonstrated  a   new  generation  of  micro-sensors  and -actuators  with increased operational capabilities and  overcoming the domination of friction over inertial forces at the micro-scale \cite{Poletkin2018}.

There are a number of techniques, which provide the implementation of  electromagnetic levitation into  a micro-actuator system and can be classified according to using materials and the sources of  the force fields  as follows: electric levitation (ELMA), magnetic levitation (MLMA) and hybrid levitation micro-actuators (HLMA). In particular, ELMA were successfully used as  linear transporters \cite{Jin1998}   and in micro-inertial sensors  \cite{Murakoshi2003,Han2012,Han2015}. MLAM can be further split into inductive (ILMA), diamagnetic (DLMA) and superconducting micro-actuators, which have found  applications in micro-bearings \cite{Coombs2005,Lu2014,PoletkinMoazenzadehMariappanEtAl2016}, micro-mirrors \cite{Shearwood1996,Xiao2018},   micro-gyroscopes \cite{Shearwood2000,Su2015a}, micro-accelerometers \cite{Garmire2007}, bistable switches \cite{Dieppedale2004} and nano-force sensors \cite{Abadie2012}.
\begin{figure}[!b]
  \centering
  \includegraphics[width=3in]{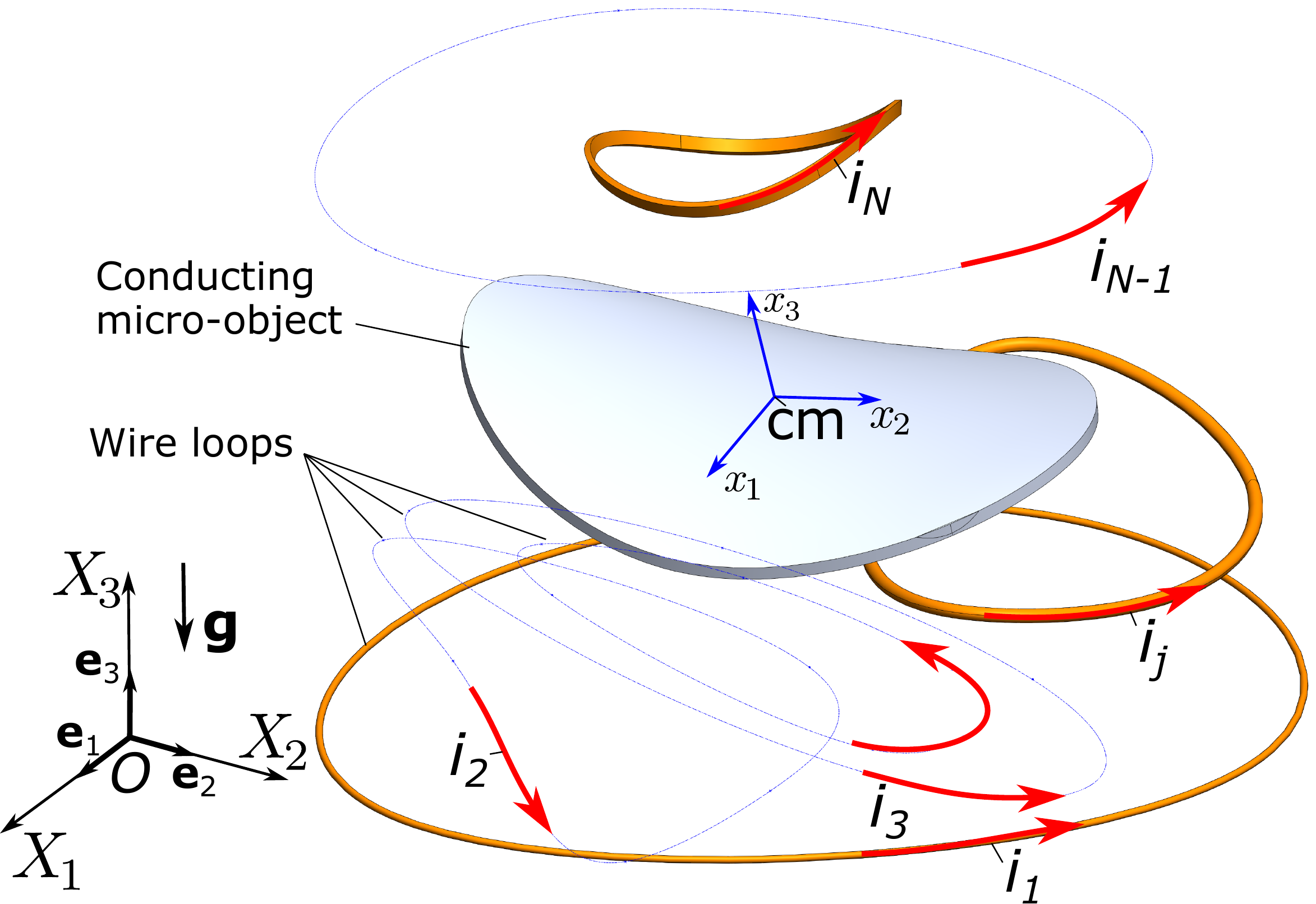}
  \caption{The generalized scheme of electromagnetic levitation system, in which a conducting micro-object is levitated by a system of arbitrary shaped wire-loops with ac currents: $X_1X_2X_3$ is the fixed coordinate frame with the corresponding unit vectors $\boldsymbol{e}_1$, $\boldsymbol{e}_2$, and $\boldsymbol{e}_3$, respectively; $\textbf{g}$ is the gravity acceleration vector directed along the $X_3$-axis; $x_1x_2x_3$  is the coordinate frame assigned to object principal axes with the origin at its centre of mass (cm); $i_j$ is the current in the $j$-wire loop.   }\label{fig:general scheme}%\vspace*{-1.5em}
\end{figure}

In HLMA different force fields are combined, for instance,   magneto- and electro-static,  variable magnetic and electro-static or
magneto-static and variable magnetic fields, which % This particularity of $\mu$-HCS
make the main difference of HLMA from  both ELAM and MLMA. In particular, capabilities of HLMA were demonstrated in applications as micro-motors \cite{Liu2008,Xu2017} and micro-accelerators \cite{Sari2014}. A wide range of different operation modes such as the linear and angular
positioning, bistable linear and angular actuations and the
adjustment of stiffness components of  a levitated micro-disc were demonstrated and experimentally studied  in the prototype  reported in \cite{Poletkin2015}. In this prototype, the  stiffness components were adjusted by changing an equilibrium
position of the inductively levitated  disc
along the vertical axis. Recently, the novel  HLMA, in which the
electrostatic forces acting on the bottom and top surfaces
of the inductively levitated micro-disc keep its equilibrium
position  and at the same time decrease a
vertical component of stiffness by means of increasing
the strength of  electrostatic field, was presented in \cite{Poletkin2017,Poletkin2018a}. A concept of this actuator for an application as a linear micro-accelerometer
was proposed in \cite{Poletkin2012}. Thus, HLMA establish a
promising direction for  further improvement in the performance of
micro-sensors and -actuators.

As seen, the electrostatic actuation is one of the main principles applied in HLMA to a passively levitated micro-object (proof mass)  for the adjustment of their static and dynamic characteristics. Simultaneously, electrostatic forces acting on the levitating micro-object  restrict its range of stable motion  by  pull-in instability. Moreover, due to the fact that the spring constant created by a magnetic contactless suspension of HLMA  has  a  nonlinear dependence on displacements, hence the pull-in phenomenon in  HLMA cannot be described and characterized  by the classic pull-in effect occurring in the spring-mass system with electrostatic actuation   \cite{ElataBamberger2006}.

Another obstacle for analysing HLMA arises due to the fact that the description of electromagnetic levitation requires the application of the Maxwell equations. Although, these equations are universally applicable, but their application even for simple designs  is
extremely difficult task \cite{Ciric1970}. Even these designs  are studied numerically by   using commercially available software, this task is still a challenge  \cite{WilliamsShearwoodMellorEtAl1997,Zhang2006,Liu2009b,Lu2012}, which is not able to cover all  aspects including stability or stable levitation and nonlinear dynamical response \cite{Laithwaite1965}.

\begin{figure}[!b]
  \centering
  \includegraphics[width=3.2in]{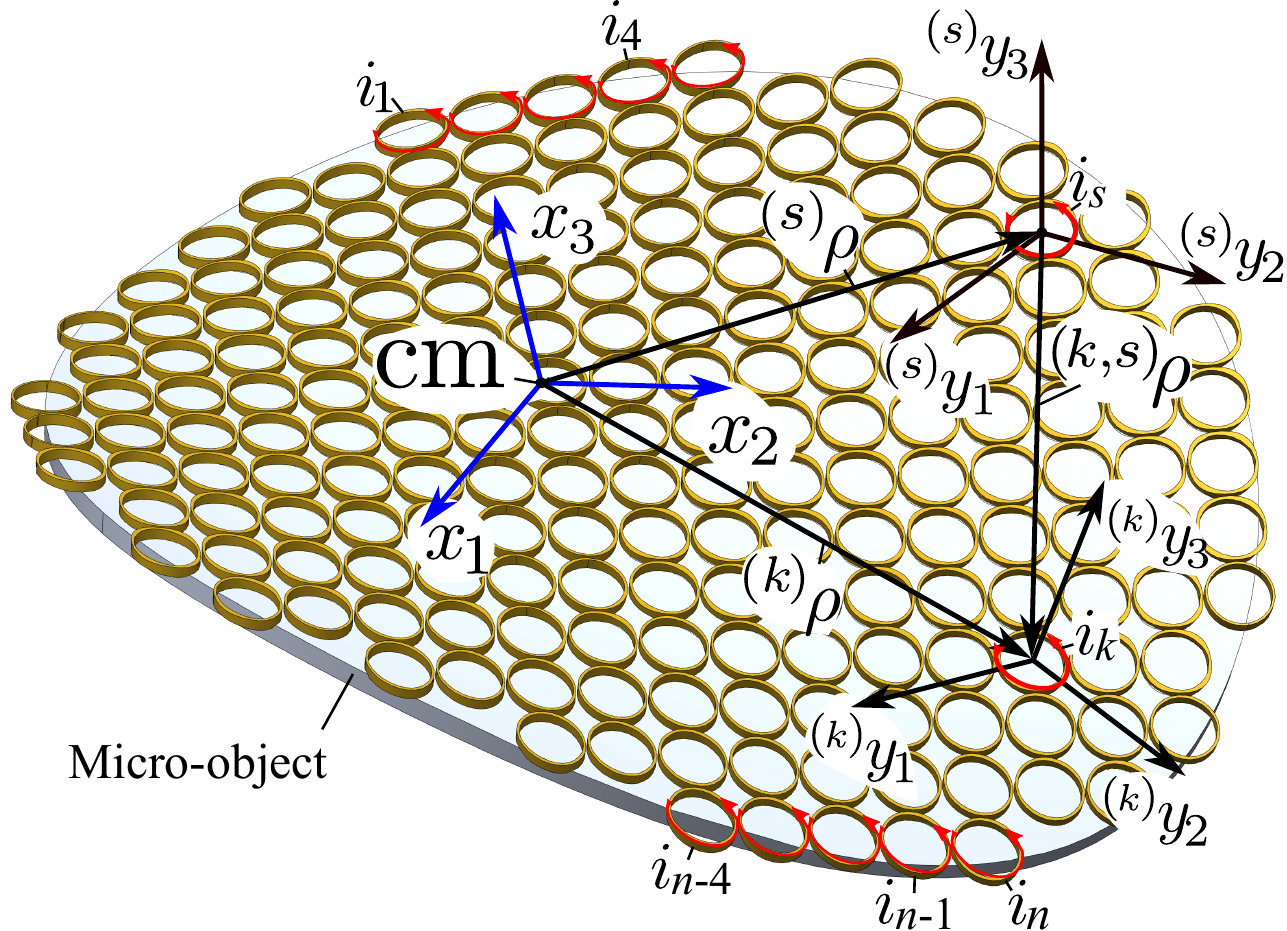}
  \caption{The conducting micro-object is meshed by finite elements of circular shape:  $\{ ^{(s)}y_l\}$ and $\{^{(k)}y_l\}$ ($l=1,2,3$) are the  coordinate frames assigned to  s- and k- circular element, respectively. }\label{fig:meshed_object}%\vspace*{-1.5em}
\end{figure}
In this work, in order to avoid having a deal with field equations, the quasi-finite element model (quasi-FEM) of HLMA based on the Lagrangian formalism is developed. The mathematical formulation of  this developed quasi-FEM becomes possible due to the recent progress in analytical calculation of mutual inductance between filament loops \cite{Babic2010,Poletkin2019}  and the technique proposed in \cite{Poletkin2014a}, which was then generalized in \cite{Poletkin2017c,Poletkin2017a}, where the induced eddy current into a levitated micro-object is considered as a finite collection of $m$-eddy current circuits.
The quasi-FEM helps us to support the mathematical reasoning of applicability of analytical modeling for, in particular, in axially symmetric designs of HLMA  to study static  pull-in instability. Although, the analytical model has some restrictions in application because of made assumptions in comparing with the quasi-FEM, which is universally applicable. However, arising some particular cases, discussing in this work, due to designs of HLMA can be  very accurately treated  by the analytical model
 presenting a solution in simple analytical forms, which
are convenient and very efficient for the practical application.
The general advantage of  both models is that they predict the static  pull-in parameters of HLMA  without needs for solving nonlinear differential equations. The results of modeling obtained by means of the
developed quasi-finite element and analytical model are verified by the
comparison with experimental results.% and in a good agreement
%with measured data.

\section{Quasi-finite element model}

Let us consider  a general design configuration  of an electromagnetic levitation system  as shown in Fig. \ref{fig:general   scheme}, which consists of $N$-arbitrary shaped wire loops and a levitated cunducting object. Each $j$-th wire loop is fed by its own ac current denoted by $i_{cj}$ with the index corresponding to  the index of the wire loop. The set of wire loops generates the alternating magnetic flux  passing through the levitated object. In turn, the eddy current is induced  within the conducting object. Force interaction between induced eddy current and  currents in wire loops provides the compensation of the gravity force acting on the conducting object along $X_3$-axis of an inertial frame \{$X_k$\} ($k=1,2,3$) and levitates then it sably  at an equilibrium position. Assuming that the levitated object is a rigid body, then  its equilibrium position can be defined through the six generalized coordinates corresponding to the three translation coordinates and three angular ones with respect to the fixed frame  \{$X_k$\} ($k=1,2,3$) with the vector base ${\boldsymbol{\underline{e}}^X}=[\boldsymbol{e}^X_1,\boldsymbol{e}^X_2,\boldsymbol{e}^X_3]^T$, where $\boldsymbol{e}^X_k$ ($k=1,2,3$) are unit vectors of ${\boldsymbol{\underline{e}}^X}$. (The notations  are adopted from the book of J. Wittenburg   \cite{Wittenburg2007}, which provide clearly distinguish between a vector, a base, and a matrix).

In order to specify these six generalized coordinates, the coordinate frame \{$x_k$\} ($k=1,2,3$) with the base ${\boldsymbol{\underline{e}}^x}$ and corresponding unit vectors $\boldsymbol{{e}}^x_k$ ($k=1,2,3$) is rigidly attached to the levitated object, in such a way, that its origin is located at the centre of mass of the object as shown in Fig. \ref{fig:general   scheme}. Also, axes of the coordinate frame \{$x_k$\} ($k=1,2,3$) coincide with principal axes of inertia of the micro-object.   The translational position of the micro-object cm with respect to the fixed frame is characterized by the vector  $\boldsymbol{q}=[q_1,q_2,q_3]^T$ and the angular one by the Brayn angles (Cardan angles) denoted as $\varphi_k$ ($k=1,2,3$). Thus, both vectors, namely, $\boldsymbol{q}$ and $\boldsymbol{\varphi}$ can be considered as  independent {\it generalized coordinates} of mechanical part of the electromagnetic levitation system.

Now let us assume that the condition of quasistationarity is hold \cite[page 7]{Martynenko1985}, \cite[page 493]{Levich1971}, hence induced eddy current within the micro-object can be represented   by $n$-electric circuits (finite elements), each of them consists of the inductor and resistor connecting in series \cite{Poletkin2017a}. Meshing the levitated micro-object by $n$-elements having the circle shape of the same radius   as shown in Fig. \ref{fig:meshed_object}, we denote $i_k$ $(k=1,2,\ldots,n)$   as the induced eddy current corresponding to the $k$-element, which can be represented as the {\it generalized velocities} of electromagnetic part of the levitation system. The linear position of $s$-circular element with respect to   the  coordinate frame \{$x_k$\} ($k=1,2,3$) is defined by the vector $^{(s)}\boldsymbol{\rho}$, but the angular position of the same element is  determined by Brayn angles, namely, $^{(s)}\boldsymbol{\phi}=[^{(s)}\phi_1,^{(s)}\phi_2, 0]^T$.

Adapting the generalized coordinates and the assumptions introduced
above, the  model can be written by using the Lagrange -
Maxwell equations as follows
\begin{equation} \label{eq:Lagrange_Maxwell}
\left\{\begin{array}{l} {\displaystyle{\frac{d}{dt} \left(\frac{\partial L}{\partial i_{k} } \right)+\frac{\partial \Psi }{\partial i_{k} } =0; \; k=1,\ldots,n};} \\
{\displaystyle{\frac{d}{dt} \left(\frac{\partial L}{\partial \dot{q_l}} \right)-\frac{\partial L}{\partial q_l} +\frac{\partial \Psi }{\partial \dot{q_l}} =F_{l} ; \;l=1,2,3;}}\\
{\displaystyle{\frac{d}{dt} \left(\frac{\partial L}{\partial \dot{\varphi_l}} \right)-\frac{\partial L}{\partial \varphi_l} +\frac{\partial \Psi }{\partial \dot{\varphi_l}} =T_{l} ; \;l=1,2,3,}} \\
 \end{array}\right.  \end{equation}
where $L=K-\Pi +W_{m}$ is the Lagrange function for the micro-object-coil system; $K=K(\dot{\boldsymbol{q}},\dot{\boldsymbol{\varphi}})$ is the kinetic energy of the system;  $\Pi=\Pi ({\boldsymbol{q}},{\boldsymbol{\varphi}})$ is the potential energy of the system; $W_{m}=W_{m}({\boldsymbol{q}},{\boldsymbol{\varphi}},i_{c1},\ldots,i_{cN},i_{1},\ldots,i_{n})$ is the energy stored in the electromagnetic field; $\Psi=\Psi (\dot{\boldsymbol{q}},\dot{\boldsymbol{\varphi}},i_{1},\ldots,i_{n})$ is the dissipation function; $F_{l}$  and $T_{l}$ ($l=1,2,3$) are the generalized force{s} and torque{s}, respectively, acting on the micro-object relative to the appropriate generalized coordinates.

The kinetic energy is
\begin{equation}
\label{eq:kinetic} {\displaystyle K=\frac{1}{2}\sum_{l=1}^3
m{\dot{q}_l}^{2} +\frac{1}{2}\sum_{l=1}^3 J_l{\dot{\omega}_l}^{2}},
\end{equation}
where $m$ is the mass of the micro-object{;} $J_l$ ($l=1,2,3$) are principal moments of inertia of the micro-object; $\omega_l=\omega_l({\boldsymbol{\varphi}},\dot{\boldsymbol{\varphi}})$ ($l=1,2,3$) are the components of the vector $\boldsymbol{\omega}$ of angular velocity of the micro-object relative to the fixed coordinate frame.

According to Fig.~\ref{fig:general scheme}, the potential energy can be defined simply as follows
\begin{equation} \label{eq:potential} \Pi =mgq_3. \end{equation}

\begin{figure*}[!t]
\centering
\includegraphics[width=5.5in]{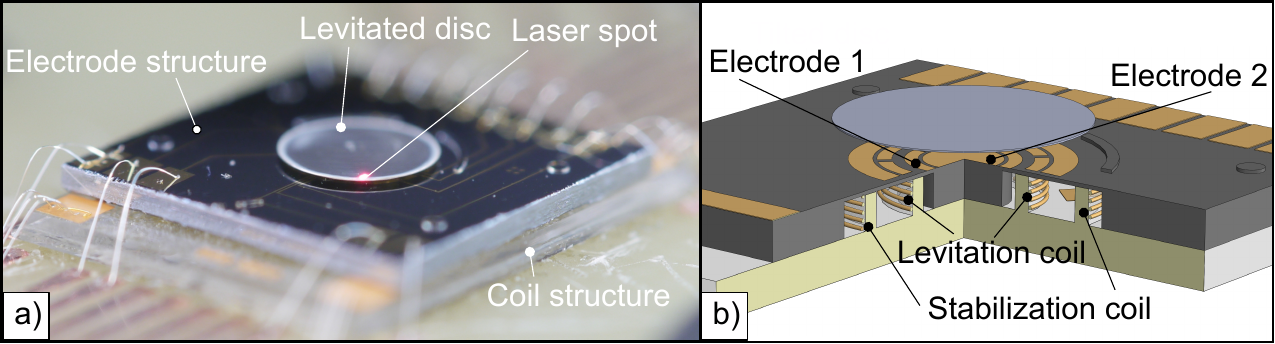}
\caption{Experimental test of pull-in actuation in the hybrid levitation micro-actuator \cite{Poletkin2015}: a) stable levitation of the disc having a diameter of 2.8 mm; b) 3D model of the device shows the location of coils and energized electrodes 1 and 2 to perform the pull-in actuation of  the levitated disc.}
\label{fig:experement}%\vspace*{-2.0em}
\end{figure*}

The dissipation function is
\begin{equation} \label{eq:dissipation}
{\displaystyle
\Psi =\frac{1}{2}\sum_{l=1}^3 \mu_l{\dot{q}_l}^{2}+\frac{1}{2}\sum_{l=1}^3 \nu_l{\dot{\varphi}_l}^{2}+\frac{1}{2}R\sum_{k=1}^n i_k^{2},
}\end{equation}
where $\mu_l$ and $\nu_l$  ($l=1,2,3$) are the damping coefficients; $R$ is the electrical resistance of the element.
 The energy stored within the electromagnetic field can be written as
\begin{equation} \label{eq:field}
\begin{split}
{\displaystyle W_{m} =\frac{1}{2}\sum_{j=1}^N\sum_{s=1}^{N}L_{js}^{c}i_{cj}i_{cs}+
\frac{1}{2}\sum_{k=1}^n\sum_{s=1}^{n}L_{ks}^{o}i_{k}i_{s}}\\
{\displaystyle +\sum_{k=1}^n\sum_{j=1}^NM_{kj}i_{k}i_{cj}} ,
\end{split}
\end{equation}
where $L_{jj}^{c}$ is the self-inductance of the $j$-wire loop; $L_{js}^{c}$, $j\neq s$ is the mutual inductance between $j$- and $s$-coils; $L_{kk}^{o}=L^{o}$ is the self-inductance of the circular  element; $L_{ks}^{o}=L_{ks}^{o}({^{(k,s)}\boldsymbol{\rho}},{^{(k,s)}\boldsymbol{\phi}})$, $k\neq s$ is the mutual inductance between $k$-  and $s$-finite circular elements (besides, $^{(k,s)}\boldsymbol{\rho}={^{(k)}\boldsymbol{\rho}}-{^{(s)}\boldsymbol{\rho}}$ and $^{(k,s)}\boldsymbol{\phi}={^{(k)}\boldsymbol{\phi}}-{^{(s)}\boldsymbol{\phi}}$, see Fig. \ref{fig:meshed_object}); $M_{kj}=M_{kj}({\boldsymbol{q}},{\boldsymbol{\varphi}})$ is the mutual inductance between  the $k$-circular element and  the $j$-wire loop.

As it was shown in works \cite{Poletkin2013,Poletkin2017a} upon assuming quasi-static behavior of the micro-object,  the induced eddy currents $i_k$ ($k=1,\ldots,n$) in circular elements can be directly expressed in terms of coil currents $i_{cj} (j=1,\ldots,N)$. Also, assuming that for each coil, the current $i_{cj}$ is a periodic signal with an amplitude of $I_{cj}$ (in general the amplitude is assumed to be a complex value) at the same frequency $f$, we can write
%\begin{equation}\label{eq:coil current}
%    i_{cj}=I_{cj}e^{\jmath f t},
%\end{equation}
$ i_{cj}=I_{cj}e^{\jmath f t},$
where $\jmath=\sqrt{-1}$. %Accounting for (\ref{eq:coil current}),
Then, the $k-$ eddy current can be
represented as $ i_k=I_ke^{\jmath f t},$
%\begin{equation}\label{eq:eddy current}
%    i_k=I_ke^{\jmath f t},
%\end{equation}
where $I_k$ is the amplitude.
 Hence, according to \cite{Poletkin2013,Poletkin2017a}  first $n$ equations of set (\ref{eq:Lagrange_Maxwell}) can be solved and the induced eddy current per circular element becomes a solution of the following set of linear equations
\begin{equation}\label{eq: the first equations of set amplitude linear}
\begin{array}{l}
  {\displaystyle{\left[L_{}^{o}+R/(\jmath f )\right] I_k+\sum_{s=1,\; s\neq k}^{n}
    L_{ks}^{o}I_s=-\sum_{j=1}^NM_{kj}I_{cj}; }} \\
   k=1,\ldots,n.
\end{array}
\end{equation}
Combining the set (\ref{eq: the first equations of set amplitude linear}) with the last six equations of (\ref{eq:Lagrange_Maxwell}) the quasi-finite element model of inductive levitation system is obtained. Thus, the obtained quasi-FEM   is the combination of finite element manner to calculate induced eddy current and the set of six differential equations describing the behavior of  mechanical part of electromagnetic levitation system.

Based on the proposed quasi-FEM the following procedure for the analysis and study of induction levitation micro-systems can be suggested.
At the beginning, the levitated micro-object is meshed by circular elements of the same radius, $R_e$, as shown in Fig. \ref{fig:meshed_object}, a value of which is defined by a number of elements, $n$.
The result of meshing becomes a list of elements $\{^{(s)}\boldsymbol{\underline{C}}=[^{(s)}\boldsymbol{\rho}, ^{(s)}\boldsymbol{\phi}]^T\}$ ($s=1,\ldots,n$) containing information about a radius vector and an angular orientation  for each element with respect to the coordinate frame \{$x_k$\} ($k=1,2,3$). Now a matrix corresponding to the left side of equation (\ref{eq: the first equations of set amplitude linear}) can be formed as follows
\begin{equation}\label{eq:matrix L}
  \underline{L}=\left[L_{}^{o}+R/(\jmath f )\right]\underline{E}+\underline{M}^{o},
\end{equation}
where $\underline{E}$ is the $(n\times n)$ unit matrix, $\underline{M}^{o}$ is the $(n\times n)$ -symmetric hollow matrix whose elements are $ L_{ks}^{o}$ ($k\neq s$). % with zero diagonal elements.
The self-inductance of the circular element is calculated by the known formula for a circular ring of circular cross-section
\begin{equation}\label{eq:self_inductance}
  L_{}^{o}=\mu_0R_e\left[\ln8/\varepsilon-7/4+\varepsilon^2/8\left(\ln8/\varepsilon+1/3\right)\right],
\end{equation}
where  $\mu_0$ is the magnetic permeability of free space, $\varepsilon=th/(2R_e)$, $th$ is the thickness of a mashed layer of micro-object.
It is recommended that the parameter $\varepsilon$ is selected to be not larger 0.1. Elements of the $\underline{M}^{o}$ matrix,  $ L_{ks}^{o}$ ($k\neq s$) can be  calculated by the formulas developed in works \cite{Babic2010,Poletkin2019}. Using the list of elements   $\{^{(s)}\boldsymbol{\underline{C}}\}$, we can estimate $L_{ks}^{o}$ as follows $ L_{ks}^{o}=L_{ks}^{o}\left(^{(k,s)}\boldsymbol{\underline{C}}\right)$, where $^{(k,s)}\boldsymbol{\underline{C}}=^{(k)}\boldsymbol{\underline{C}}-^{(s)}\boldsymbol{\underline{C}}$.

Then, we assign to  each coil the coordinate frame \{$^{(j)}z_k$\} ($k=1,2,3$) with the corresponding base $^{(j)}\boldsymbol{\underline{e}}^z$. The linear and angular position of \{$^{(j)}z_k$\} ($k=1,2,3$) with respect to the fixed  frame \{$X_k$\} ($k=1,2,3$) is defined by the radius vector $^{(j)}\boldsymbol{{r}_c^X}$ and the Brayn angles $^{(j)}\boldsymbol{\phi_c}$  $(j=1,\ldots,N)$, respectively. Knowing a projection of the $j$-coil filament loop on  the $^{(k)}y_1$-$^{(k)}y_2$ plane of the $k$- circular element,
the Kalantarov-Zeitlin method \cite{Poletkin2019} can be used in order to calculate the mutual inductance. Hence, the $(n\times N)$  matrix $\underline{M}_c$ consisting of elements of the mutual inductance $M_{kj}$ can be obtained.
The induced eddy current in each circular element is a solution of the following matrix equation
\begin{equation}\label{eq:matrix eq}
  \underline{I}= \underline{L}^{-1}\underline{M}_c\underline{I}_c,
\end{equation}
where $\underline{I}$ is  the  $(n\times 1)$ matrix of eddy currents and $\underline{I}_c=[I_{c1} I_{c2} \ldots I_{cN}]^T$ is the given  $(N\times 1)$ matrix of currents in coils.

Finally, substituting the  eddy current  $\underline{I}$ into the stored electromagnetic energy (\ref{eq:field}) the ponderomotive forces acting on the levitated object can be found as
the first derivative of the stored electromagnetic energy with  respect to  the { generalized coordinates} of mechanical part. Thus, in a matrix form we can write as
\begin{equation} \label{eq:ponderomotive forces}
\begin{array}{l}
{\displaystyle{F_{l}=\frac{\partial W_m}{\partial q_l}= \underline{I}^T\frac{\partial \underline{M}_c}{\partial q_l}\underline{I}_c; \;l=1,2,3;}}\\
{\displaystyle{T_{l}=\frac{\partial W_m}{\partial \varphi_l}=\underline{I}^T\frac{\partial \underline{M}_c}{\partial \varphi_l}\underline{I}_c; \;l=1,2,3.}} \\
 \end{array}
  \end{equation}

\section{Hybrid Levitation Micro-actuators}
\label{sec:design}
\begin{figure}[!t]
\centering
\includegraphics[width=3.4in]{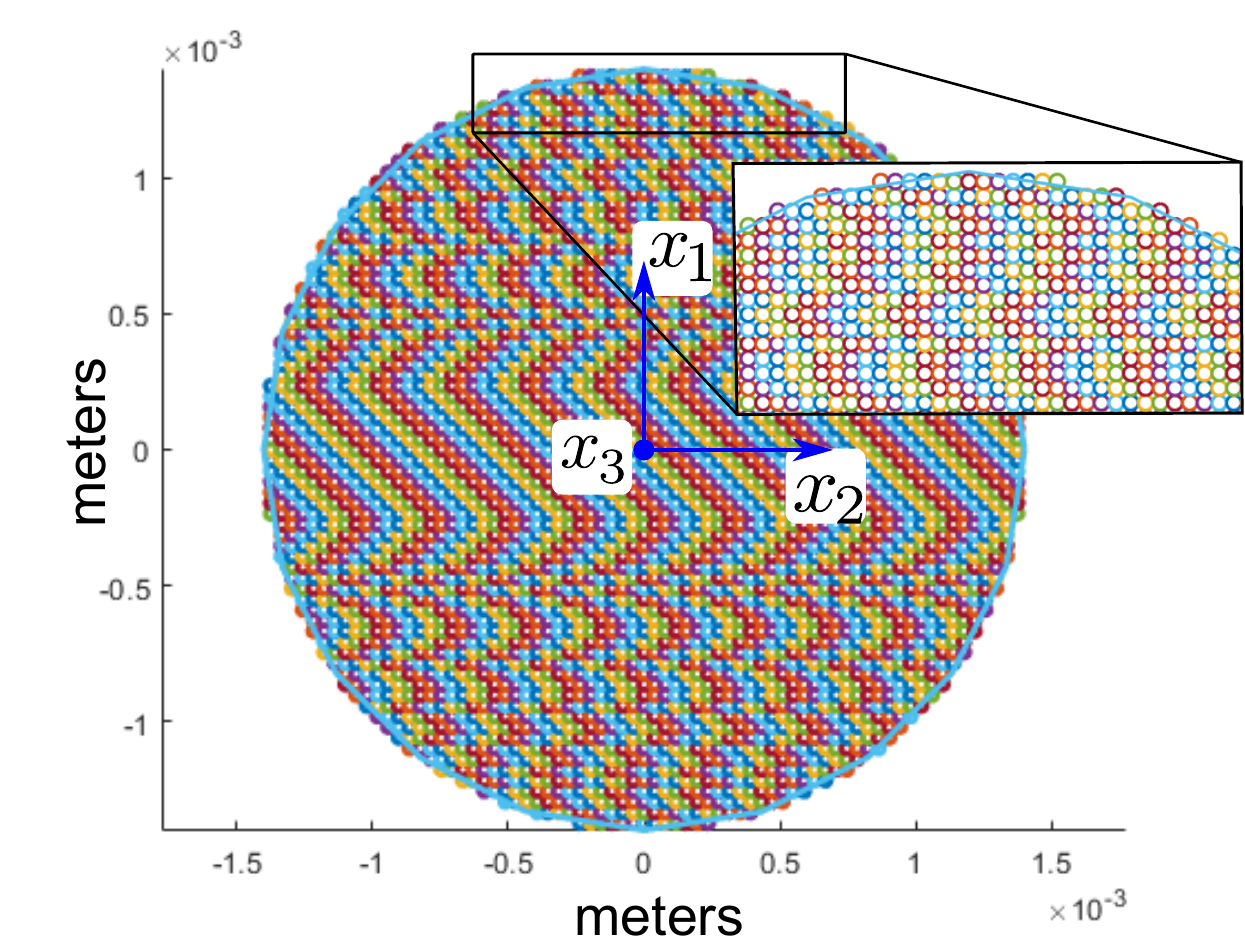}
\caption{Disc of a diameter of \SI{2.8}{\milli\metre} is meshed by 3993 circular elements.}
\label{fig:mesh}%\vspace*{-2.0em}
\end{figure}

\begin{figure}[!b]
\centering
\includegraphics[width=3.0in]{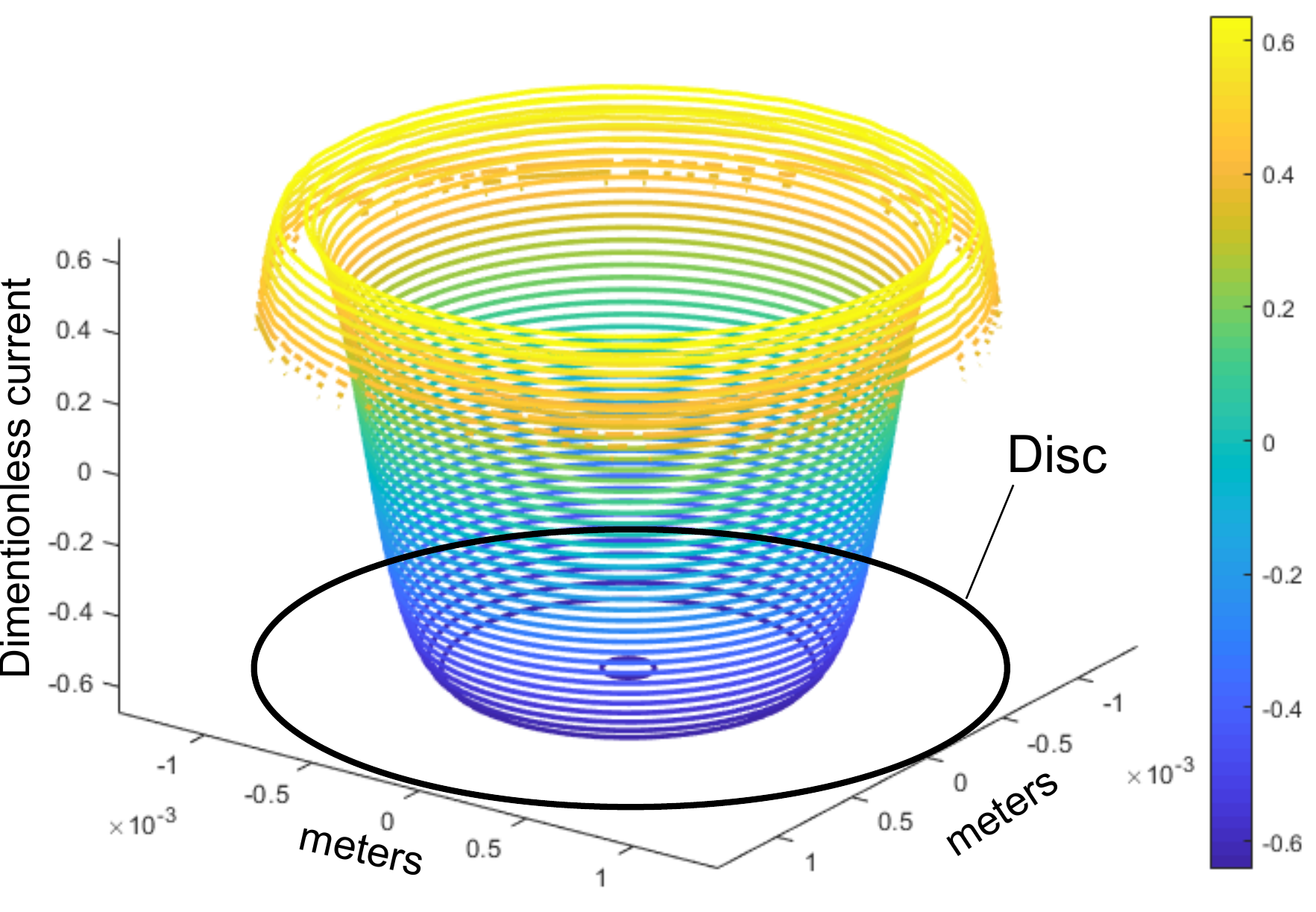}
\caption{The distribution of the eddy current in mesh circular elements.}
\label{fig:eddy current per element}%\vspace*{-2.0em}
\end{figure}
Let us consider one of the  designs of hybrid levitation micro-actuators base on the two coil structure for stable levitation  and electrode structure to generate  electrostatic force acting on the bottom surface of the levitated micro-disc. In particular, such the design was implemented into the prototype  of  HLMA, which is shown in Fig. \ref{fig:experement}a) and was studied experimentally in works \cite{Poletkin2015,PoletkinLuWallrabeEtAl2015}. Two coaxial solenoidal microcoils of the coil structure was fabricated on the pyrex substrate and were excited by ac current having a   frequency of 10 MHz.
The diameters
of the two coils shown in Fig. \ref{fig:experement}b), namely, levitation and stabilization one were \SIlist{2000;3800}{\micro\metre}%2000 and 3800 $\mu$m
, respectively. The inner (levitation) coil had
20 windings and the outer (stabilisation) coil had 12 windings of a gold wire with a diameter of \SI{25}{\micro\metre}. %25 $\mu$m.
The coil structure was covered by the electrode structure fabricated on a Si wafer. According to the design, the electrodes were located above  the coil post at a height of \SI{25}{\micro\metre}. % 40  $\mu$m.
In order to carry out the pull-in actuation of the disc, electrodes 1 and 2, having the same area, $A_e$, of $8.0\cdot10^{-7}$ m$^2$ as shown in Fig. \ref{fig:experement}b), were energized in a way that the disc was moved toward the electrode surface. The detailed results of the experimental studies and measurements of the pull-in actuation such as the pull-in voltage and displacement had been presented and discussed in \cite{Poletkin2015,PoletkinLuWallrabeEtAl2015}.  In this paper, the pull-in mechanism in the considered design of HLMA is comprehensively studying theoretically by means of analytical modelling and numerical simulation based on the developed quasi-FEM.

\begin{figure}[!t]
\centering
\includegraphics[width=2.9in]{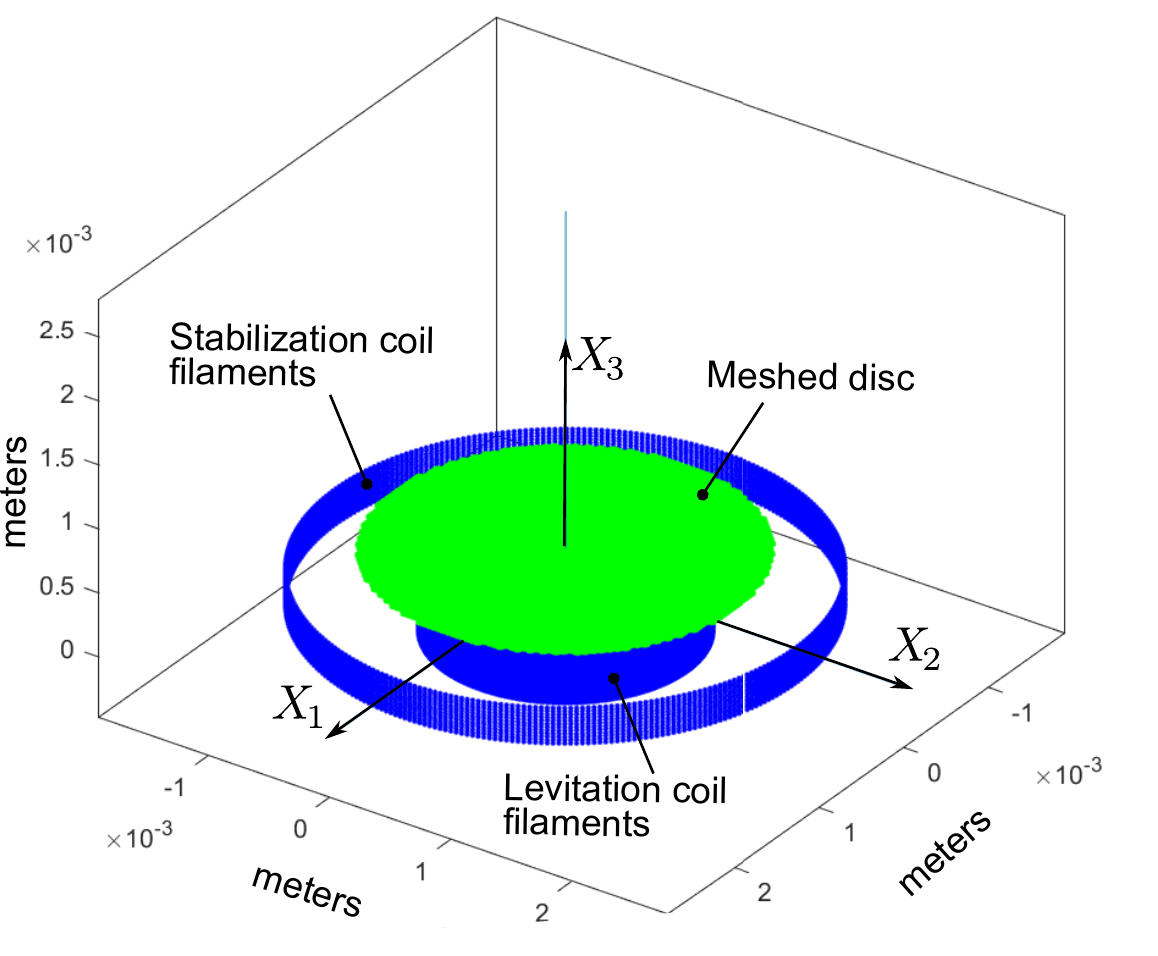}
\caption{3D geometrical scheme of HLMA for eddy current simulation: \{$X_k$\} ($k=1,2,3$) is the fixed coordinate frame.}
\label{fig:3Dscheme}%\vspace*{-2.0em}
\end{figure}

\begin{figure}[!b]
\centering
\includegraphics[width=2.7in]{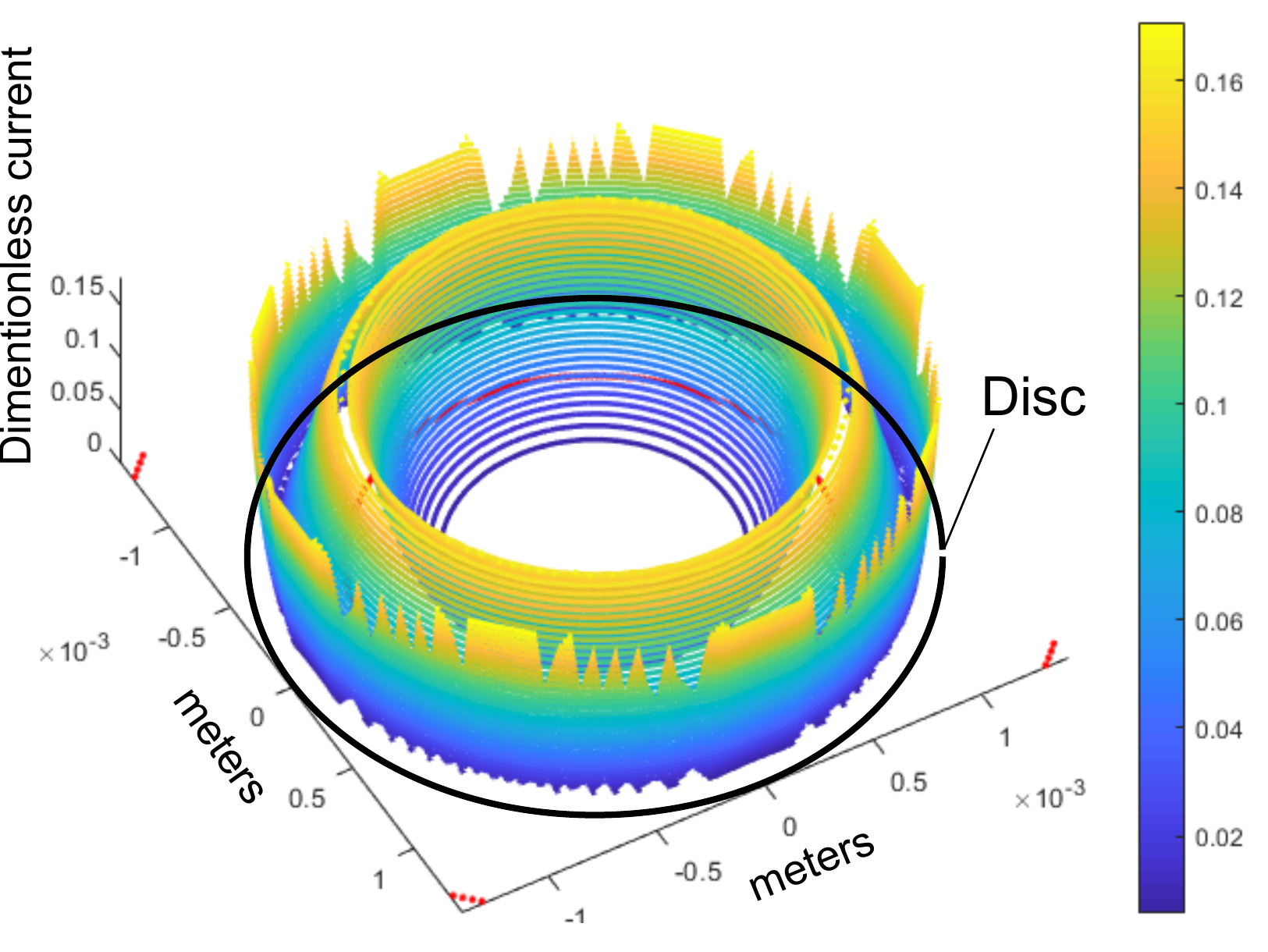}
\caption{The distribution of  magnitudes of eddy current with respect to unit vectors of $\boldsymbol{{e}}^x_1$ and $\boldsymbol{{e}}^x_2$ of  the base  ${\boldsymbol{\underline{e}}^x}$.}
\label{fig:eddy current magnitude}%\vspace*{-2.0em}
\end{figure}

\begin{figure*}[t]
  \centering
  \includegraphics[width=5in]{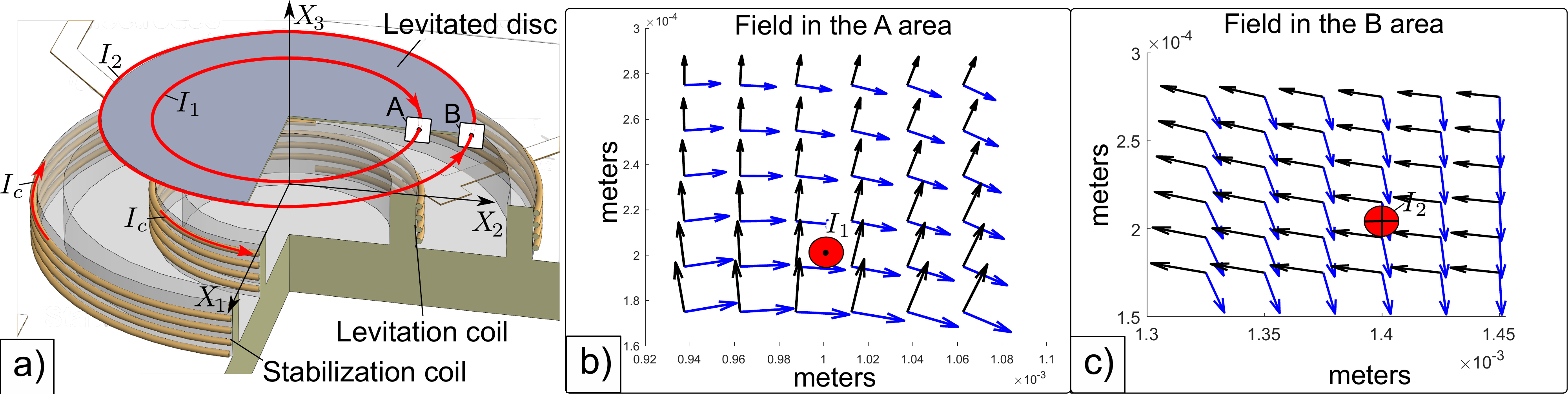}
  \caption{Modelling of linear pull-in actuation: a) scheme of HLMA based on the approximation of  induced eddy current in the disc by two eddy current circuits with  $I_1$ and $I_2$ currents corresponding to its maximum magnitudes; b) the map of magnetic field (blue arrows) and its corresponded gradient (black arrows)  within the A area build on the $X_3$-$X_2$ plane around the $I_1$ circuit (see Fig. \ref{fig:scheme_field}a)); c) the map of magnetic field and its corresponded gradient within the B area build on the $X_3$-$X_2$ plane around the $I_2$ circuit (see Fig. \ref{fig:scheme_field}a)).    }\label{fig:scheme_field}%\vspace*{-1.5em}
\end{figure*}

\section{Eddy current simulation}
\label{sec:simulation}
In this section  a distribution of induced eddy current within a disc having a diameter of \SI{2.8}{\milli\metre}  levitated at height, $h_l$,  of \SI{200}{\micro\metre}  by the coil structure corresponding to the design of HLMA described above in Sec. \ref{sec:design} is simulated by the developed quasi-FEM. According to the proposed procedure, the disc is homogenously meshed by 3993 circular elements.  A result of meshing is a map of the location of  elements      with respect to the coordinate frame \{$x_k$\} ($k=1,2,3$), the origin of which is placed at the centre of the disc, as shown in Fig \ref{fig:mesh}. Each element crosses its neighboring element only at one point. Depending on the  location of  neighboring element at the top, right, bottom and left side, this point can be placed on the element perimeter at an angle of \SIlist{0;90;180;270}{\degree} subtended at the center of the circular element, respectively.    Elements  are numbering from left to right in each line.  Because of the plane shape of the levitated micro-object, vectors of the list of elements $\{^{(s)}\boldsymbol{\underline{C}}\}$ have the following structures, namely,    $^{(s)}\boldsymbol{\rho}=[^{(s)}x_1,^{(s)}x_2,0]^T$ and $^{(s)}\boldsymbol{\phi}=[0,0,0]^T$ ($s=1,\ldots,n$). Knowing $\{^{(s)}\boldsymbol{\underline{C}}\}$, the $\underline{L}$ matrix can be calculated by Eq. (\ref{eq:matrix L}).

3D geometry of two micro-coils is approximated by a series of circular filaments. Hence, depending on the number of windings,  the levitation coil is replaced by 20 circular filaments, while the stabilization coil by 12 circular filaments. Thus, the total number of circular filaments, $N$, is 32. Assigning the origin of the fixed frame \{$X_k$\} ($k=1,2,3$) to the centre of the circular filament corresponding  to the first top  winding of the levitation coil, the linear position of the circular filaments of levitation coil  can be defined as $^{(j)}\boldsymbol{{r}_c}=[0,0,(j-1)\cdot p]^T$, $(j=1,\ldots,20)$, where $p$ is the pitch equaling to  \SI{25}{\micro\metre}.  The same is applicable for stabilization coil, $^{(j)}\boldsymbol{{r}_c}=[0,0,(j-21)\cdot p]^T$,  with the difference that the index $j$ is varied from 21 by 32.  For both coils,  the Brayn angle of each circular filament is  $^{(j)}\boldsymbol{\phi_c}=[0,0,0]^T$,   $(j=1,\ldots,32)$. Accounting for the values of  diameters of levitation and stabilization coils,  3D geometrical scheme of HLMA for eddy current simulation can be build as shown in Fig. \ref{fig:3Dscheme}.

The position of the coordinate frame \{$x_k$\} ($k=1,2,3$) with respect to the fixed frame  \{$X_k$\} ($k=1,2,3$) is defined by the radius vector  $\boldsymbol{{r}}_{cm}=[0,0,h_l]^T$.
Then, the position of the $s$-mesh element  with respect to the coordinate frame \{$^{(j)}z_k$\} ($k=1,2,3$) assigned to the $j$-coil filament can be found as  $^{(s,j)}\boldsymbol{{r}}=\boldsymbol{{r}}_{cm}+{^{(s)}\boldsymbol{\rho}}-{^{(j)}\boldsymbol{{r}}}_{c}$
or in a matrix form as
\begin{equation}\label{eq:r(s,j)}
  ^{(s,j)}\underline{{r}}^z={^{(j)}\underline{A}^{zX}}{\underline{r}}_{cm}^X+{^{(j)}\underline{A}}^{zx}{^{(s)}\underline{\rho}^x}-^{(j)}\underline{A}^{zX}{^{(j)}\underline{{r}}_{c}^X},
\end{equation}
where $^{(j)}\underline{A}^{zX}={^{(j)}\underline{A}^{zX}}\left({^{(j)}\boldsymbol{\phi_c}}\right)={^{(j)}\boldsymbol{\underline{e}}^z}\cdot{\boldsymbol{\underline{e}}^X}$
and
$^{(j)}\underline{A}^{zx}={^{(j)}\underline{A}^{zX}}\left({^{(j)}\boldsymbol{\phi_c}}\right)\underline{A}^{Xx}(\boldsymbol{\varphi})={^{(j)}\boldsymbol{\underline{e}}^z}\cdot{\boldsymbol{\underline{e}}^x}$ are the direction cosine matrices. Because all angles are zero, hence $^{(j)}\underline{A}^{zx}={^{(j)}\underline{A}^{zX}}=\underline{E}$, where $\underline{E}$ is the ($3\times3$) unit matrix.
Since the coils are represented by the circular filaments and using the radius vector $^{(s,j)}\boldsymbol{{r}}$, the mutual inductance  between the $j$- coil and $s$-meshed element  can be calculated directly by the formula presented in \cite{Poletkin2019}. Thereby, the $\underline{M}_c$  matrix can be formed.
It is convenient to present the result of calculation in the dimensionless form. For this reason, the dimensionless currents in the levitation coil and stabilization one  are introduced by dividing currents on the amplitude of the current in the levitation coil.  Since the amplitudes of the current in both coil are the same.
Hence,  the input current in the levitation coil filaments are to be  one, while in the stabilization coil filaments to be minus one. Now, the induced eddy current in  dimensionless values can be calculated by Eq. (\ref{eq:matrix eq}).

In order to illustratively present the calculation result, the obtained ($3993\times 1$)  eddy current  matrix, $\underline{I}$, is transformed into the ($71\times 71$)  2D matrix, $\underline{I}$. Data in this ($71\times 71$)  2D matrix are allocated   similarly to the  structure corresponding to Fig. \ref{fig:mesh}.
Then, the distribution along the disc surface of induced eddy current in mesh circular elements   are shown in Fig. \ref{fig:eddy current per element}. The analysis of Fig. \ref{fig:eddy current per element} shows that in a central area of the disc, corresponding to the area of the circular cross-section of the levitation coil, the eddy current has the negative sign (it means that the direction of  induced eddy current flow is opposite to the flow direction  in the levitation coil) due to the significant contribution of the ac magnetic field generated by   this coil. While, outside of this area, a sign becomes positive due to the ac  magnetic field of the stabilization coil.

Now let us present the obtained  result in the vector form through unit vectors $\boldsymbol{{e}}^x_1$ and $\boldsymbol{{e}}^x_2$ of the base  ${\boldsymbol{\underline{e}}^x}$. Taking  the numerical gradient of  the ($71\times 71$)  2D matrix, $\underline{I}$ with respect to the rows and columns, the components in the form of the ($71\times 71$)  2D matrixs of $\underline{I}_1$  and  $\underline{I}_2$ relative to the unit vectors $\boldsymbol{{e}}^x_1$ and $\boldsymbol{{e}}^x_2$ are calculated, respectively. Then, the ($71\times 71$)  2D matrix of  magnitudes of the eddy current for each mesh point  is estimated by $\sqrt{{\underline{I}_1}^2+{\underline{I}_2}^2}$. The result of estimation is shown in Fig. \ref{fig:eddy current magnitude}. Fig. \ref{fig:eddy current magnitude} depicts that maximum magnitudes of eddy current are concentrated along the edge of the disc and in its central part along the circle having the same  diameter as the levitation coil. This result is similar to one obtained by Lu in work \cite{Lu2012}, where the induced eddy current in a  disc levitated by two coils with the  similar design   was simulated by COMSOL software. Both results provide the reason and applicability of the approximation based on a two eddy current circuits, for the analysis of HLMAs with an axially symmetric design, which was proposed in \cite{Poletkin2017a} and successfully used for the study of their stability and dynamics.

\section{Analytical model of static linear pull-in actuation}

In this section, the approximation of induced eddy current within the disc based on  two eddy current circuits is applied to model  the linear pull-in actuation in HLMA with the design under consideration. Scheme for modelling is shown in Fig. \ref{fig:scheme_field}a). Due to particularities of the HLMA design, a radius of the $I_1$ circuit is equal to \SI{1}{\milli\meter}, while   a radius of the $I_2$ circuit is restricted by the radius of disc equaling to \SI{1.4}{\milli\meter}.  Let us examine the field distribution around these two circles of eddy current circuits. Since the design is axially symmetric, it is enough to consider some areas such as denoted by  $A$  and $B$ as shown in Fig. \ref{fig:scheme_field}a)  cutting, for instance, on the   $X_3$-$X_2$ plane  and crossing the $I_1$  and $I_2$ circuit, respectively. Mapping the field in the $A$  and $B$   area  shows that the force acting on  the $I_1$ circuit determined by the corresponded gradient of the field is directed vertically and lifts up the disc, while the force acting on  the $I_2$ circuit has  almost horizontal direction and pushes the disc toward the center  as presented in Fig.\ref{fig:scheme_field}b) and c), respectively. Noting that  the names of coils as levitation and stabilization coil reflect their functionalities coming form the field analysis conducted above.
It can be assumed that for a diameter of disc equaling to  \SI{2.8}{\milli\meter} the influence  of the $I_2$ circuit on the linear pull-in actuation is small and can be neglected.
Thus, modelling of the linear pull-in actuation is reduced to the force interaction between the currents in the levitation coil and the $I_1$   circuit.

Then, the behavior of the disc along the $X_3$ axis can be described as follows \cite{Poletkin2018b}:
\begin{equation}\label{eq:model I}
  m \frac{d^2q_3}{dt^2}+mg+\frac{I_c^2}{L^o}\frac{dM}{dq_3}M+\frac{A_0}{4}\frac{U^2}{(h+q_3)^2}=0,
\end{equation}
where $U$ is the applied voltage to the electrodes (namely, electrode 1 and 2  shown in Fig. \ref{fig:experement}b)), which have the same area of $A_e$, $A_0=\varepsilon_0A_e$, $\varepsilon_0$ is the permeability of free space, $h$ is the space between the electrode surface and the origin of  the coordinate frame \{$x_k$\} ($k=1,2,3$) measured along the $X_3$ axis.
For this particular case,  the mutual inductance, $M$, can be defined by the Maxwell formula \cite[page 6]{Rosa1912} such as
\begin{equation}\label{eq:M}
  \begin{array}{l}
    {\displaystyle
                k^2=\frac{4R_l^2}{4R_l^2+(h_l+y)^2}};\\
    {\displaystyle
    M=\mu_0R_l\left[\left(\frac{2}{k}-k\right)K(k)-\frac{2}{k}E(k)\right]},
  \end{array}
\end{equation}
where $\mu_0$ is the magnetic permeability of free space, $R_l$ is the radius of levitation coil, $K$ and $E$ are complete elliptic integrals of the first and second kind, respectively \cite{Dwight1961}.
\begin{figure}[!t]
\centering
\includegraphics[width=2.3in]{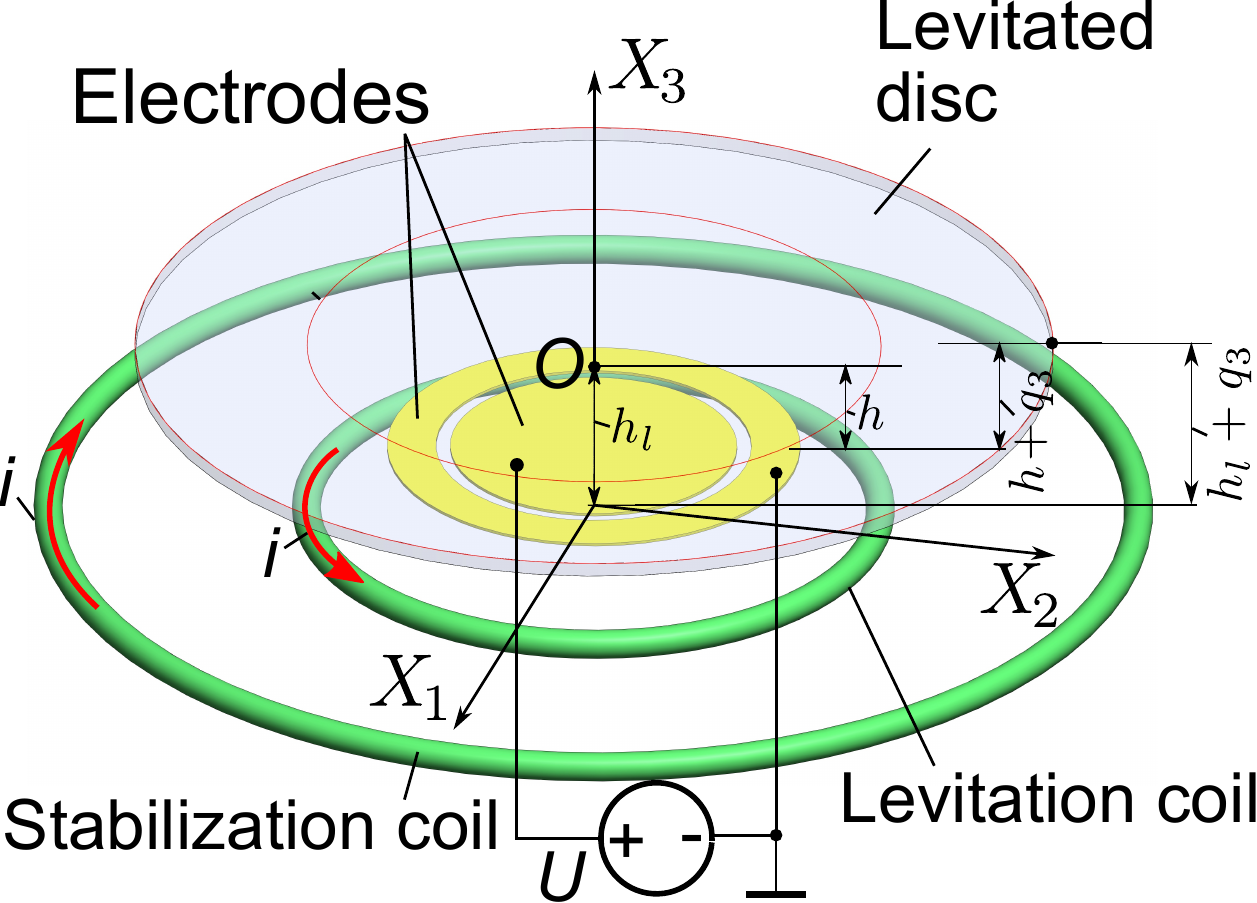}
\caption{The simplest scheme  for preliminary analysis of obtained models: $U$ is the potential applied to electrodes, $h$ is the space between an electrode plane
and equilibrium point of proof mass, $h_l$ is the levitation height between a plane of coils and equilibrium point of the disc.}
\label{fig:scheme}%\vspace*{-2.0em}
\end{figure}

 For  further analysis, model (\ref{eq:model I}) is presented in the dimensionless form  as follows
 \begin{equation}\label{eq:model II elliptic dimensionless}
\begin{array}{l}
 {\displaystyle
\frac{d^2\lambda}{d\tau^2}+1-\eta\left[\left(\frac{2}{k}-k\right)K(k)-\frac{2}{k}E(k)\right]\frac{2}{k^2}}\\
  {\displaystyle
  \times\left[\frac{2-k^2}{2(1-k^2)}E(k)-K(k)\right]\cdot\frac{\kappa\xi^2(1+\kappa\lambda)}{(1+\xi^2(1+\kappa\lambda)^2)^{3/2}}}\\
  {\displaystyle
  +\frac{\beta}{(1+\lambda)^2}=0,}
  \end{array}
\end{equation}
where $\tau=\sqrt{g/h}t$, $\lambda=q_3/h$, $\eta=I^2a^2/(mghL^o)$, $\beta=A_0U^2/(4mgh^2)$, $\kappa=h/h_l$, $a=R_l\mu_0$ and $\xi=h_l/(2R_l)$.

From Eq. (\ref{eq:model II elliptic dimensionless}), the static pull-in model is
 \begin{equation}\label{eq:analytic pull-in dimensionless}
\begin{array}{l}
 {\displaystyle
\beta=(1+\lambda)^2\left(-1+\eta\left[\left(\frac{2}{k}-k\right)K(k)-\frac{2}{k}E(k)\right]\frac{2}{k^2}\right.}\\
  {\displaystyle
 \left. \times\left[\frac{2-k^2}{2(1-k^2)}E(k)-K(k)\right]\cdot\frac{\kappa\xi^2(1+\kappa\lambda)}{(1+\xi^2(1+\kappa\lambda)^2)^{3/2}}\right),}\\
   \end{array}
\end{equation}
where the $\eta$ constant can be defined at equilibrium state of the system, when $\lambda$ and $\beta$ are zero, as follows
 \begin{equation}\label{eq:eta constant}
\begin{array}{l}
 {\displaystyle
\eta=\left(\left[\left(\frac{2}{k_0}-k_0\right)K(k_0)-\frac{2}{k_0}E(k_0)\right]\frac{2}{k_0^2}\right.}\\
  {\displaystyle
 \left. \times\left[\frac{2-k_0^2}{2(1-k_0^2)}E(k_0)-K(k_0)\right]\cdot\frac{\kappa\xi^2}{(1+\xi^2)^{3/2}}\right)^{-1},}\\
   \end{array}
\end{equation}
and $k_0^2=1/(1+\xi^2)$.

\begin{figure}[!t]
\centering
\includegraphics[width=2.5in]{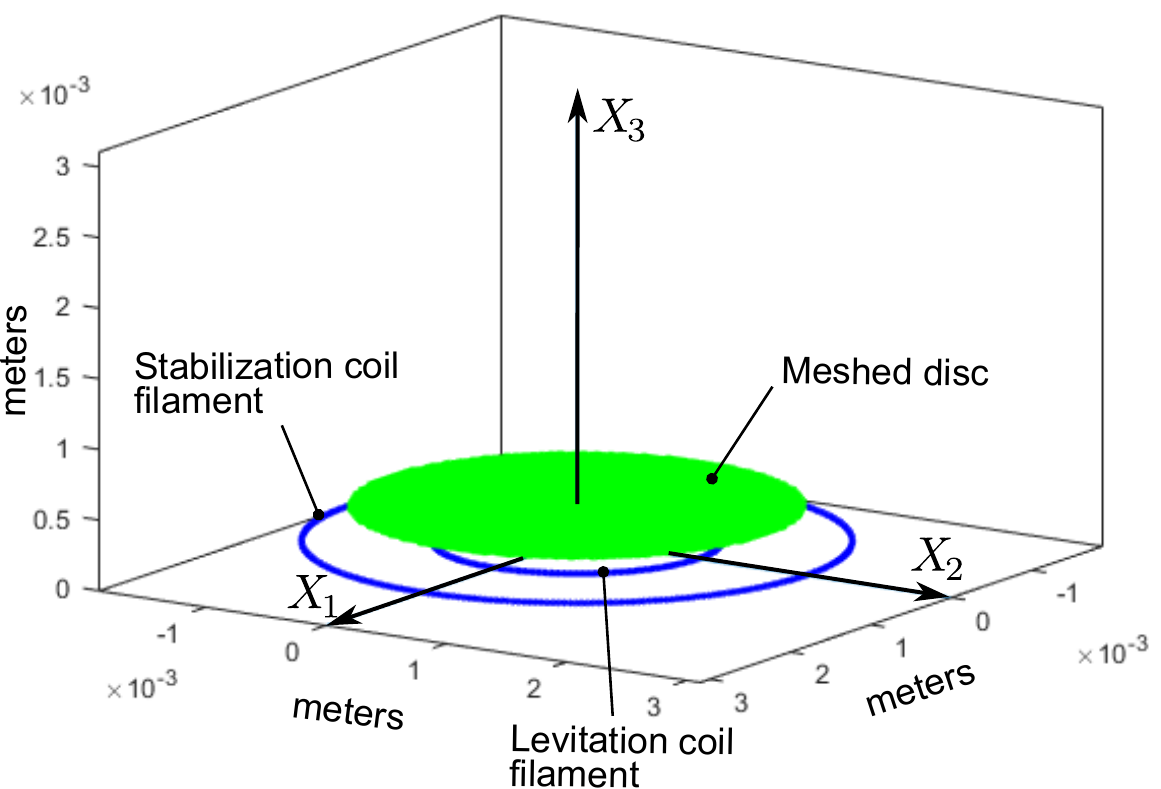}
\caption{The 3D scheme for simulation:  the \SI{3.1}{\milli\meter} diameter disc is meshed by 3993 circular elements.}
\label{fig:3Dscheme_plane}%\vspace*{-2.0em}
\end{figure}

\section{Quasi-FEM of static linear pull-in actuation}

The quasi-FEM model of the static linear pull-in has a similar form to (\ref{eq:model I}). The difference arises due to the fact  that the magnetic interaction between the disc and coils  along the  $X_3$ axis is defined by the force $F_3$ from Eq. (\ref{eq:ponderomotive forces}). Hence, taking into account this fact, the quasi-FEM model becomes
\begin{equation}\label{eq:model quasi}
  m \frac{d^2q_3}{dt^2}+mg+\underline{I}^T\frac{\partial \underline{M}_c}{\partial q_3}\underline{I}_c+\frac{A_0}{4}\frac{U^2}{(h+q_3)^2}=0.
\end{equation}
Now, we present Eq. (\ref{eq:model quasi}) in dimensionless from:
\begin{equation}\label{eq:model quasi dimensionless}
\begin{array}{l}
 {\displaystyle
\frac{d^2\lambda}{d\tau^2}+1+\eta_0F_m(\lambda) +\frac{\beta}{(1+\lambda)^2}=0};\\
  {\displaystyle
  F_m(\lambda)=\sum_{s=1}^{n}\sum_{j=1}^{N}\eta_{sj}\frac{\partial \overline{M}_{sj}(\overline{x}_1,\overline{x}_2,(1+\lambda\kappa)\chi)}{\partial \lambda}},
  \end{array}
\end{equation}
where $\eta_0=\mu_0 I_{c1}^2\sqrt{R_{c1}R_e}/(mgR_e)$, $R_{c1}$ is the radius of the first winding of the  levitation coil, $\eta_{sj}=\overline{I}_s\overline{I}_{cj}\sqrt{\overline{R}_{cj}}\Big/\chi$, $\overline{I}_s=I_s/I_{c1}$ and $\overline{I}_{cj}=I_{cj}/I_{c1}$ are the dimensionless currents,
$\overline{R}_{cj}=R_{cj}/R_{c1}$, $\chi=h_l/R_e$ is the scaling factor, $\partial \overline{M}_{sj}/{\partial \lambda}$ is the derivative of dimensionless mutual inductance with respect to $\lambda$ (its definition is shown in Appendix \ref{app}), $\overline{x}_1=x_1/R_e$ and $\overline{x}_2=x_2/R_e$ are the dimensionless coordinates. Noting that $x_1$ and $x_2$ are defined by Eq. (\ref{eq:r(s,j)}).

The static pull-in model based on quasi-FEM (\ref{eq:model quasi dimensionless}) is
\begin{equation}\label{eq:quasi pull-in dimensionless}
 {\displaystyle
\beta=-(1+\lambda)^2\big(1+\eta_0F_m(\lambda)\big)},
 \end{equation}
where similar to (\ref{eq:eta constant}) the $\eta_0$ constant is also defined at equilibrium state  as follows
 \begin{equation}\label{eq:eta zero constant}
 {\displaystyle
\eta_0=-1/F_m(0).}
\end{equation}

\begin{figure}[!t]
  \centering
  \includegraphics[width=3.6in]{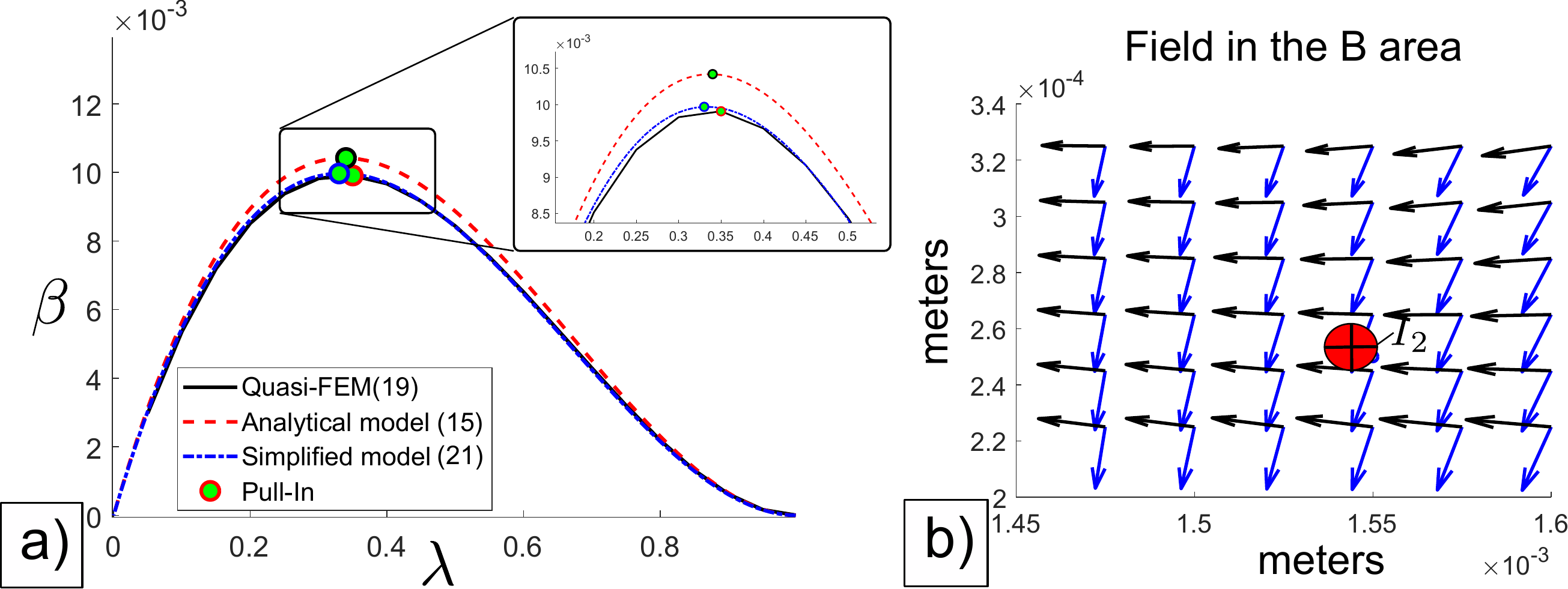}
  \caption{Pull-in actuation of the  \SI{3.1}{\milli\meter} diameter disc: a) equilibrium curve of the square voltage vs displacement (absolute value of $\lambda$); b) the map of the magnetic field (blue arrows) and its corresponded gradient (black arrows) within the B area build on the $X_3$-$X_2$ plane around the $I_2$ circuit.    }\label{fig:pull_in_1_55mm_field}%\vspace*{-1.5em}
\end{figure}

\begin{figure}[b]
  \centering
  \includegraphics[width=3.4in]{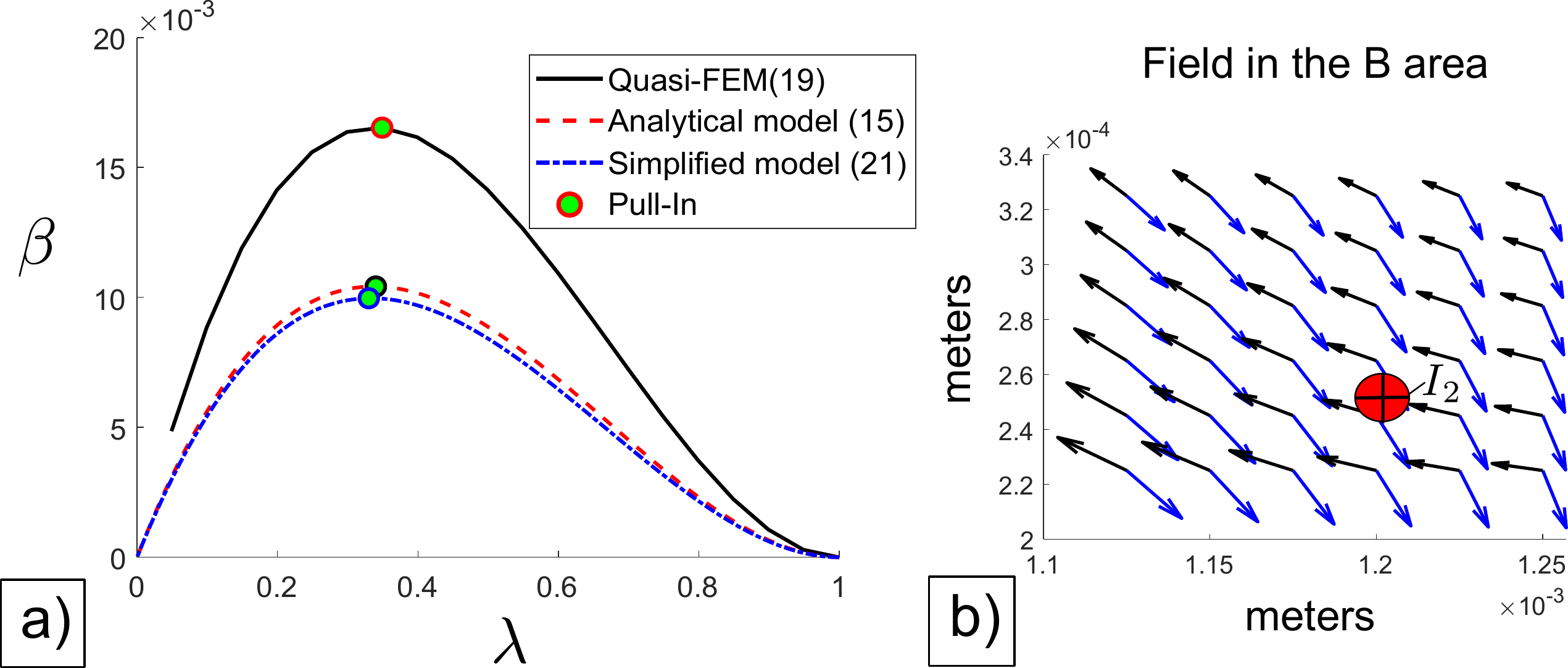}
  \caption{Pull-in actuation of the  \SI{2.4}{\milli\meter} diameter disc: a) equilibrium curve of the square voltage vs displacement (absolute value of $\lambda$); b) the map of the magnetic field (blue arrows) and its corresponded gradient (black arrows) within the B area build on the $X_3$-$X_2$ plane around the $I_2$ circuit.    }\label{fig:pull_in_1_2mm_field}%\vspace*{-1.5em}
\end{figure}
\section{Preliminary  analysis of developed models }

It is obvious that if the magnetic field gradient in the B area around the $I_2$ circuit of eddy current within the disc and corresponding force is directed almost horizontally (see Fig. \ref{fig:scheme_field}c)), then    the estimation of pull-in parameters by means of the analytical model (\ref{eq:analytic pull-in dimensionless}) becomes close to the exact calculation performed by the quasi-FEM (\ref{eq:quasi pull-in dimensionless}).   This fact indicates that the application of the analytical model requires  the  knowledge  about the  gradient of the magnetic field  in the B area of a particular design under consideration.
%This point  is discussed in the current section.

On the other hand, due to design particularities of HLMA there are some particular cases, which   can be immediately treated by  the  model (\ref{eq:analytic pull-in dimensionless})  presenting a solution in simple analytical equations. In turn, these simple equations are convenient for the practical application. For instance, let us consider such a particular case, which takes place when dimensionless parameter $\kappa$ is small.  From physical point of view, it means that the redistribution between  energy  stored within the electrical field of capacitors and energy  stored within magnetic field of coils and levitated disc occurs  by  changing the location of electrodes along the $X_3$ axis and, in particular, the electrodes are located closer to the levitated disc. This fact leads to the possibility of a  linearization of the magnetic force in   the analytical model (\ref{eq:analytic pull-in dimensionless}). Thus, we can write the following simple model of static pull-in actuation as
\begin{equation}\label{eq:simple pull-in dimensionless}
 {\displaystyle
\beta=-\frac{\ln(4/\xi)-1}{\ln(4/\xi)-2}\kappa\lambda(1+\lambda)^2}.
 \end{equation}
From the later model, the pull-in parameters can be estimated to be
\begin{equation}\label{eq:pull-in parameters}
  \lambda_p=-\frac{1}{3}, \;\;\; \beta_p=\frac{\ln(4/\xi)-1}{\ln(4/\xi)-2}\kappa\frac{4}{27}.
\end{equation}

\begin{figure}[t!]
  \centering
  \includegraphics[width=3.2in]{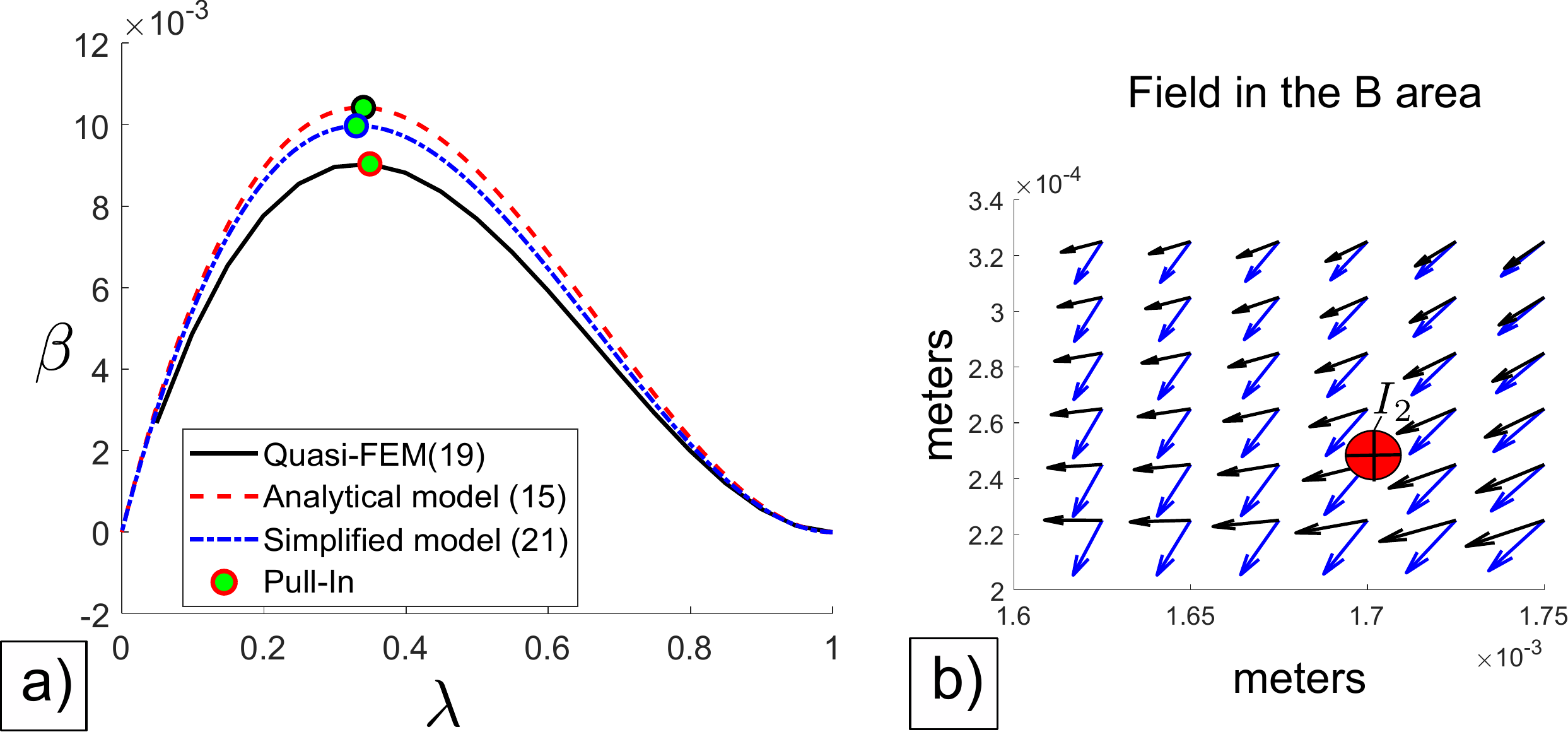}
  \caption{Pull-in actuation of the  \SI{3.2}{\milli\meter} diameter disc: a) equilibrium curve of the square voltage vs displacement (absolute value of $\lambda$); b) the map of the magnetic field (blue arrows) and its corresponded gradient (black arrows) within the B area build on the $X_3$-$X_2$ plane around the $I_2$ circuit.    }\label{fig:pull_in_1_7mm_field}%\vspace*{-1.5em}
\end{figure}

\begin{table}[b!]
\caption{\label{tab:disc 1.55mm}Pull-in parameters for the  \SI{3.1}{\milli\meter} diameter disc.}
%\vspace*{-1.0em}
\begin{center}
\begin{tabular}{lccccc}
\hline
\multirow{2}{*}{Model}
&\multicolumn{3}{c}{Pull-in parameter} &\multicolumn{2}{c}{Relative Error}   \\
& $\lambda_p$ & $\beta_p$ &$\sqrt{\beta_p}$& $\Delta\lambda_p$&$\Delta\sqrt{\beta_p}$ \\
\hline
Quasi-FEM, (\ref{eq:quasi pull-in dimensionless}) & 0.34 &0.0099 &0.0995& --& --\\
Analytical model, (\ref{eq:analytic pull-in dimensionless}) & 0.34&0.0104 &0.102& 0 & 0.025 \\
Simplified model, (\ref{eq:simple pull-in dimensionless}) & 0.3333 &0.01 & 0.1&0.0474& 0.005 \\
\hline
\end{tabular}%\vspace*{-3.0em}
\end{center}
\end{table}

Now let us apply the obtained models, namely, (\ref{eq:analytic pull-in dimensionless}), (\ref{eq:quasi pull-in dimensionless}) and (\ref{eq:simple pull-in dimensionless}) to the design shown in Fig. \ref{fig:scheme} and consists of two circular plane coils having radii of \SI{1.0}{\milli\meter} and \SI{1.9}{\milli\meter} for levitation and stabilization coil, respectively. The disc is levitated at height, $h_l$, of \SI{250}{\micro\meter}, while the electrodes are placed  at a point  measured   from the equilibrium point $O$ of the disc along the $X_3$ axis on a  distance, $h$, of \SI{10}{\micro\meter}. Hence, the dimensionless parameters of the design become $\kappa=0.04$  and $\xi=0.125$. Then, according to Eq. (\ref{eq:pull-in parameters}) the square dimensionless pull-in voltage can be calculated and becomes $\beta_p=1.4930\frac{4}{27}=0.01$.
The modelling is performed for a disc with the following radii: \SI{1.2}{}, \SI{1.55}{} and \SI{1.7}{\milli\meter}.  Fig. \ref{fig:3Dscheme_plane} depicts the 3D scheme for simulation. The disc is meshed by 3993 circular elements.

The three equilibrium curves of pull-in actuation for a disc  having a diameter of \SI{1.55}{\milli\meter} calculated by models (\ref{eq:analytic pull-in dimensionless}), (\ref{eq:quasi pull-in dimensionless}) and (\ref{eq:simple pull-in dimensionless}) are shown in Fig. \ref{fig:pull_in_1_55mm_field}a). Results of estimation of pull-in parameters are summed up in Table \ref{tab:disc 1.55mm}. The simulation shows that the pull-in displacement (in absolute value) is $\lambda_p=0.34$  and the pull-in voltage is $\sqrt{\beta_p}=0.0995$.  The relative errors of estimation of pull-in parameters by means of the analytical models do not exceed \SI{5}{\%} (please see Table \ref{tab:disc 1.55mm}). Noting that the map of gradient of the magnetic field in the B area is directed almost horizontally as shown in Fig.\ref{fig:pull_in_1_55mm_field}b). Hence,  the contribution of electromagnetic force due to interaction between the magnetic field and  the $I_2$ circuit to the pull-in actuation is small.

\begin{figure}[!t]
  \centering
  \includegraphics[width=3.5in]{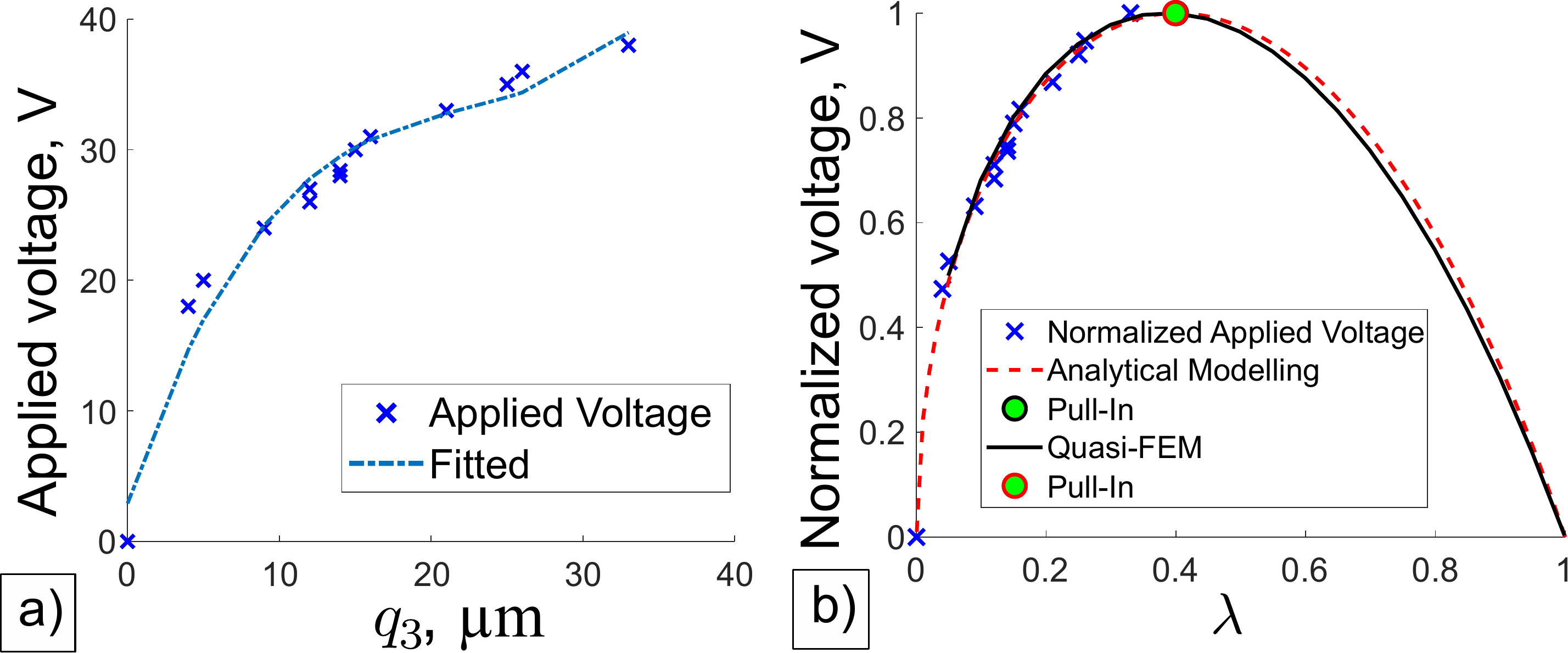}
  \caption{Pull-in actuation of disc having a  \SI{2.4}{\milli\meter} diameter: a) experimental measurement of applied voltage vs linear displacement; b) normalized voltage  vs dimensionless displacement: measurement data together with equilibrium curves generated by quasi-FEM and analytical one.    }\label{fig:exper_disc 2_4mm}%\vspace*{-1.5em}
\end{figure}
\begin{figure}[!b]
  \centering
  \includegraphics[width=2.5in]{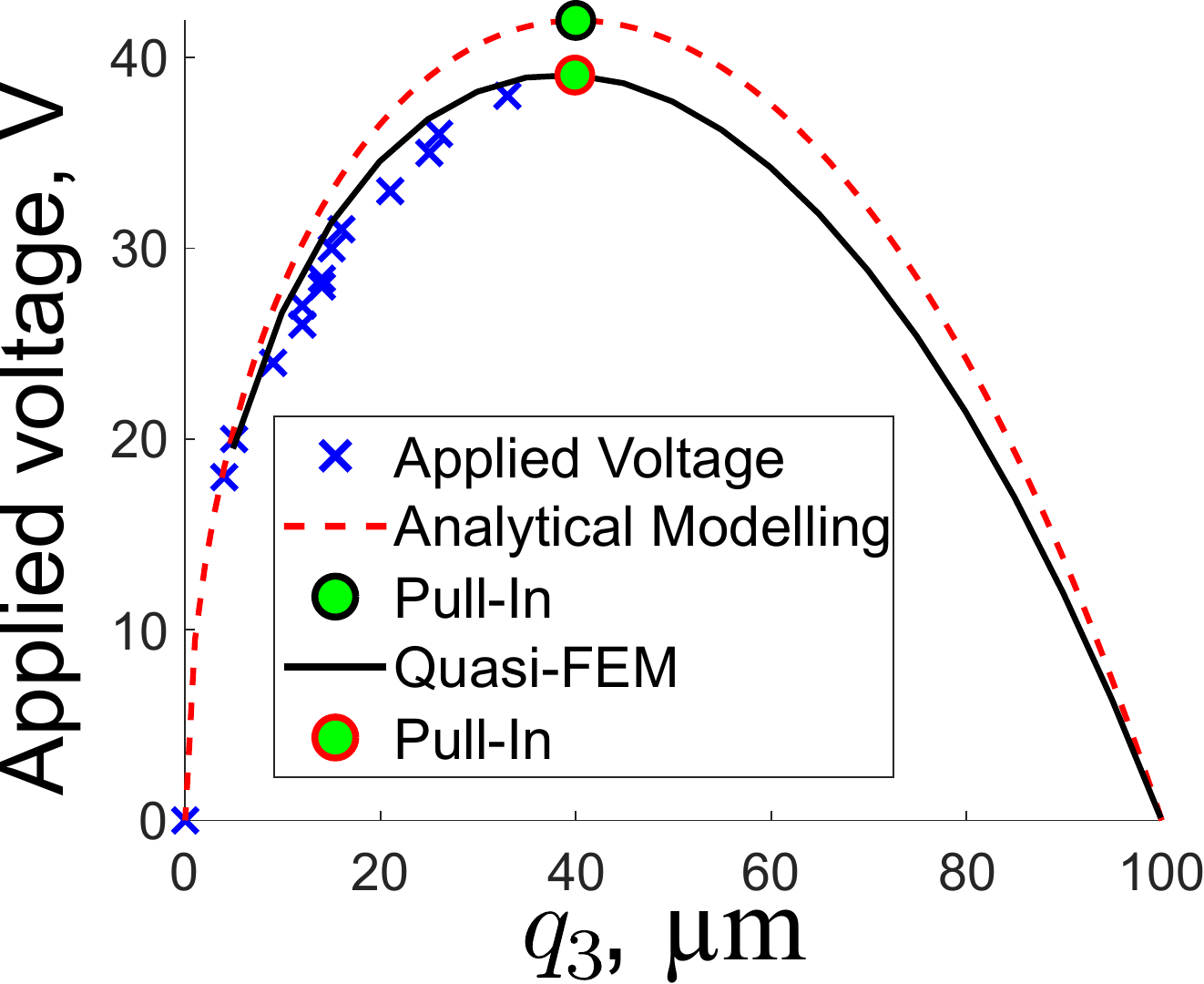}
  \caption{ Two equilibrium curves calculated by quasi-FEM Eq. (\ref{eq:model quasi})  and analytical model Eq. (\ref{eq:model I}) of applied voltage vs displacement of the disc having a  \SI{2.4}{\milli\meter} diameter in comparing with experimental data.    }\label{fig:exper_disc 2_4mm absolute}%\vspace*{-1.5em}
\end{figure}

Fig. \ref{fig:pull_in_1_2mm_field}a) shows the equilibrium curve of  pull-in actuation for a disc  having a diameter of \SI{2.4}{\milli\meter}, which is  simulated by quasi-FEM (\ref{eq:quasi pull-in dimensionless}).
Since the analytical models  (\ref{eq:analytic pull-in dimensionless}) and (\ref{eq:simple pull-in dimensionless}) are independent of a radius of the levitated disc, the results of modelling  are  the same as presented in Fig.\ref{fig:pull_in_1_55mm_field}a) and  Table \ref{tab:disc 1.55mm}.
The simulation predicts the following values of the pull-in parameters such as
\begin{equation}\label{eq:pull-in parameters of disc 1.2mm}
  \lambda_p=0.34, \;\;\; \beta_p=0.01653, \;\;\; \sqrt{\beta_p}=0.1286,
\end{equation}
where the dimensionless displacement $\lambda_p$ is given in absolute value. Although, the pull-in displacement has the same value as in the previous example, the main difference appears in the estimation of the pull-in voltage, which is increased due to the contribution of the electromagnetic force exerted on the $I_2$ circuit as shown in Fig. \ref{fig:pull_in_1_2mm_field}b). As a result, the relative error of the calculation of pull-in voltage by means of the analytical models is also increased drastically to   \SI{30}{\%}.

In the case of the  \SI{3.4}{\milli\meter} diameter disc, the equilibrium curve simulated by quasi-FEM (\ref{eq:quasi pull-in dimensionless}) is shown in Fig. \ref{fig:pull_in_1_7mm_field}a). The simulation predicts the following pull-in parameters such as
\begin{equation}\label{eq:pull-in parameters of disc 1.2mm}
  \lambda_p=0.34, \;\;\; \beta_p=0.009,\;\;\; \sqrt{\beta_p}=0.0949.
\end{equation}
In this case, the calculation of pull-in voltage by means of analytical models shows that  the relative error is around \SI{5}{\%}.
As seen, the value of the pull-in voltage is decreased  due to  the electromagnetic force acting on the $I_2$ circuit, which has a vertical component  directed against   the levitation force as shown in Fig. \ref{fig:pull_in_1_7mm_field}b).

Noting that in all three cases considered above, the map of the magnetic field and the corresponded gradient in the A area is similar to the map shown in  Fig. \ref{fig:scheme_field}b).

\begin{figure}[!t]
  \centering
  \includegraphics[width=3.5in]{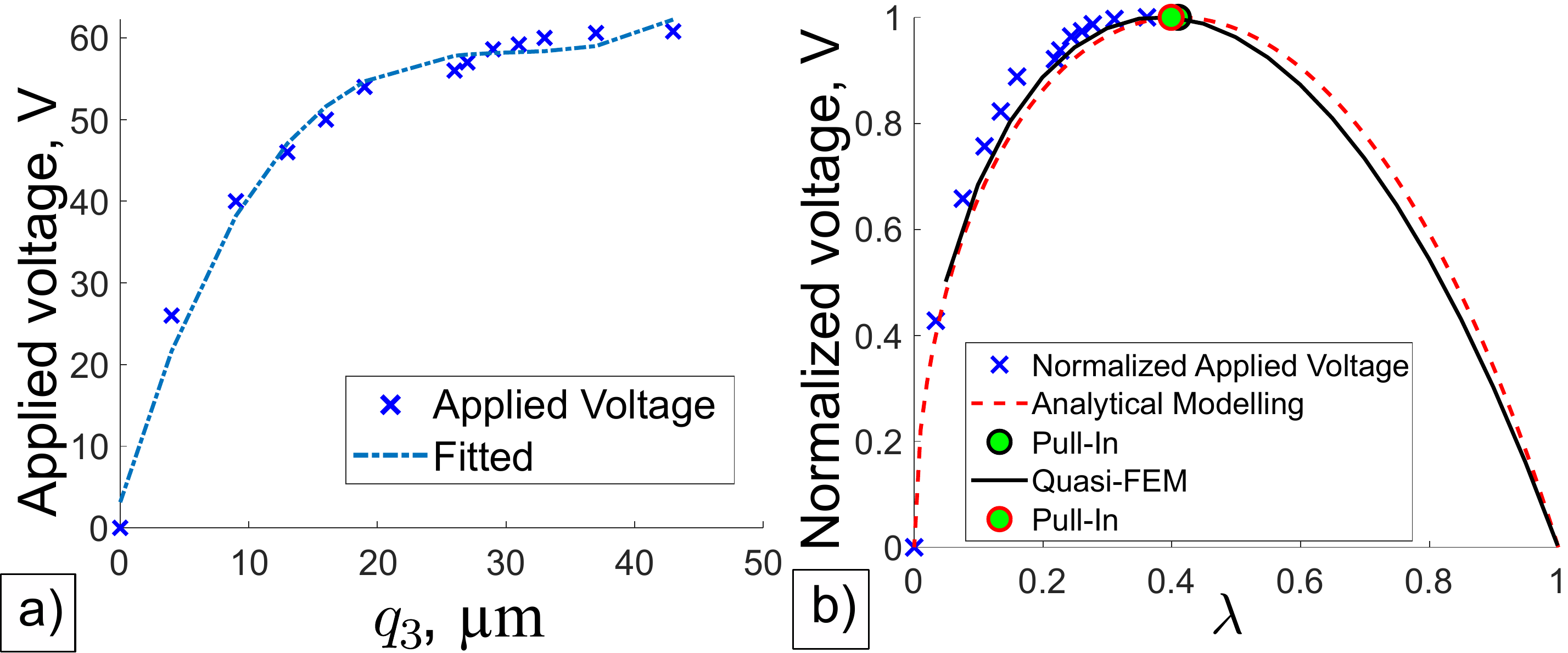}
  \caption{Pull-in actuation of disc having a  \SI{2.8}{\milli\meter} diameter: a) experimental measurement of applied voltage vs linear displacement \cite{Poletkin2015}; b) normalized voltage  vs dimensionless displacement: measurement data together with equilibrium curves generated by quasi-FEM and analytical one.   }\label{fig:exper_disc 2_8mm}%\vspace*{-1.5em}
\end{figure}
\begin{figure}[!b]
  \centering
  \includegraphics[width=2.5in]{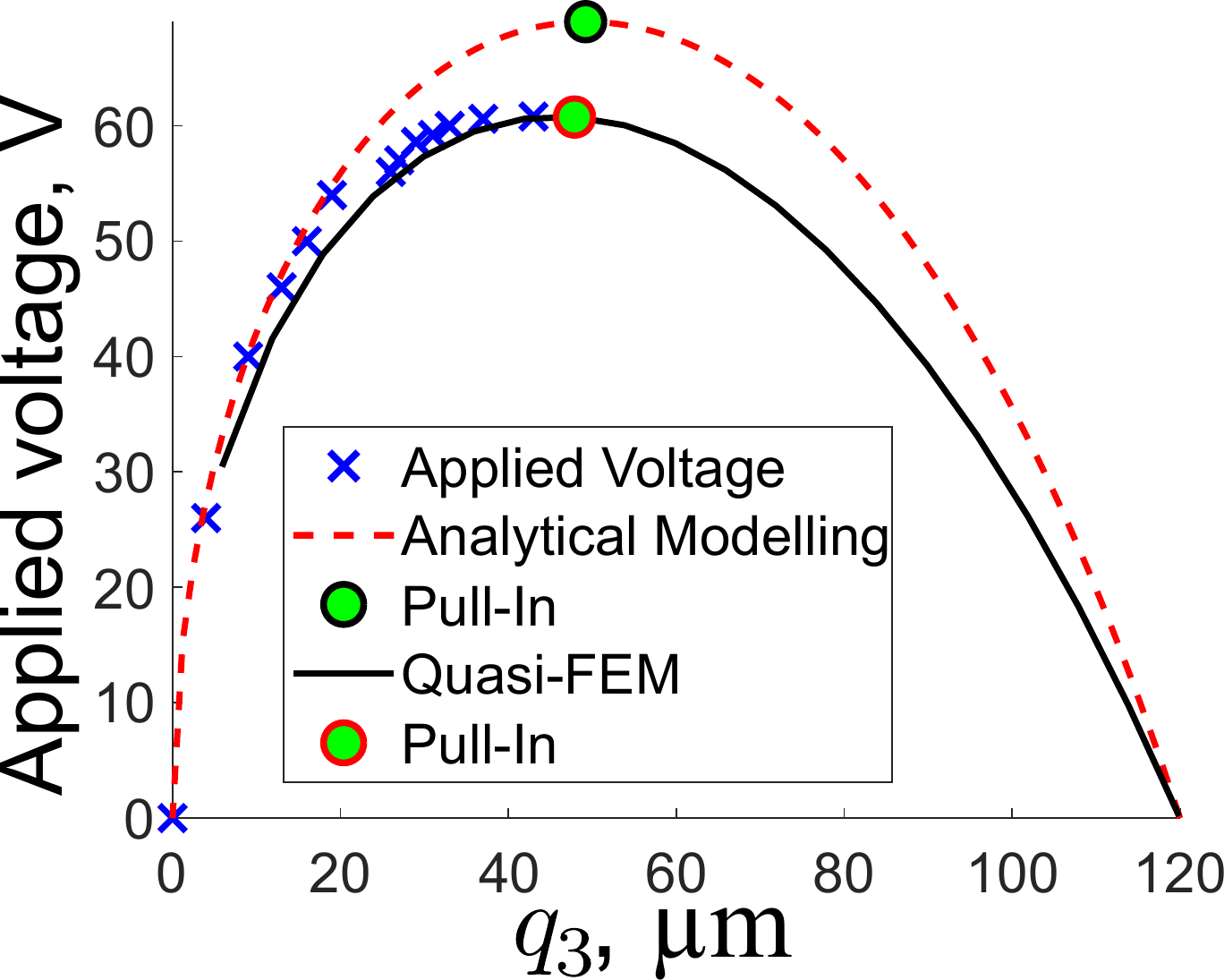}
  \caption{ Two equilibrium curves calculated by quasi-FEM Eq. (\ref{eq:model quasi})  and analytical model Eq. (\ref{eq:model I}) of applied voltage vs displacement of the disc having a  \SI{2.8}{\milli\meter} diameter in comparing with experimental data.    }\label{fig:exper_disc 2_8mm absolute}%\vspace*{-1.5em}
\end{figure}

\section{Comparison with experiment}
Now let us compare the results of simulation and modelling generated by the quasi-FEM  and analytical model, respectively, with experimental data reported in \cite{Poletkin2015,PoletkinLuWallrabeEtAl2015} for linear pull-in actuation performed by two discs having  diameters of \SI{2.8}{} and \SI{3.2}{\milli\meter} in the prototype of HLMA presented in Sec. \ref{sec:design}. Also, we added new experimental data collected in the same prototype for the \SI{3.2}{\milli\meter} diameter disc and a lighter disc of a \SI{2.4}{\milli\meter} diameter.

\begin{figure}[!t]
  \centering
  \includegraphics[width=3.5in]{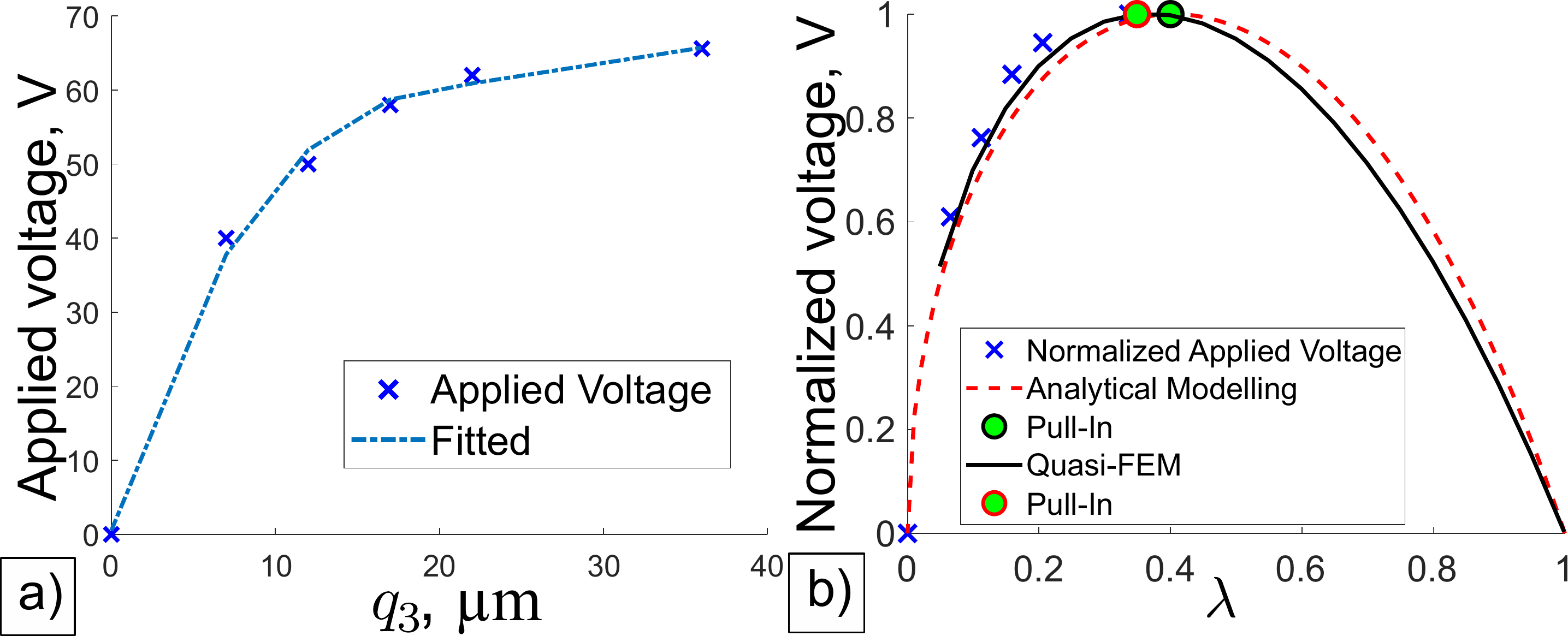}
  \caption{Pull-in actuation of disc having a  \SI{3.2}{\milli\meter} diameter and levitated at \SI{107}{\micro\metre}: a) experimental measurement of applied voltage vs linear displacement \cite{PoletkinLuWallrabeEtAl2015}; b) normalized voltage  vs dimensionless displacement: measurement data together with equilibrium curves generated by quasi-FEM and analytical one. }\label{fig:exper_disc 3_2mm_107}%\vspace*{-1.5em}
\end{figure}
\begin{figure}[!b]
  \centering
  \includegraphics[width=2.5in]{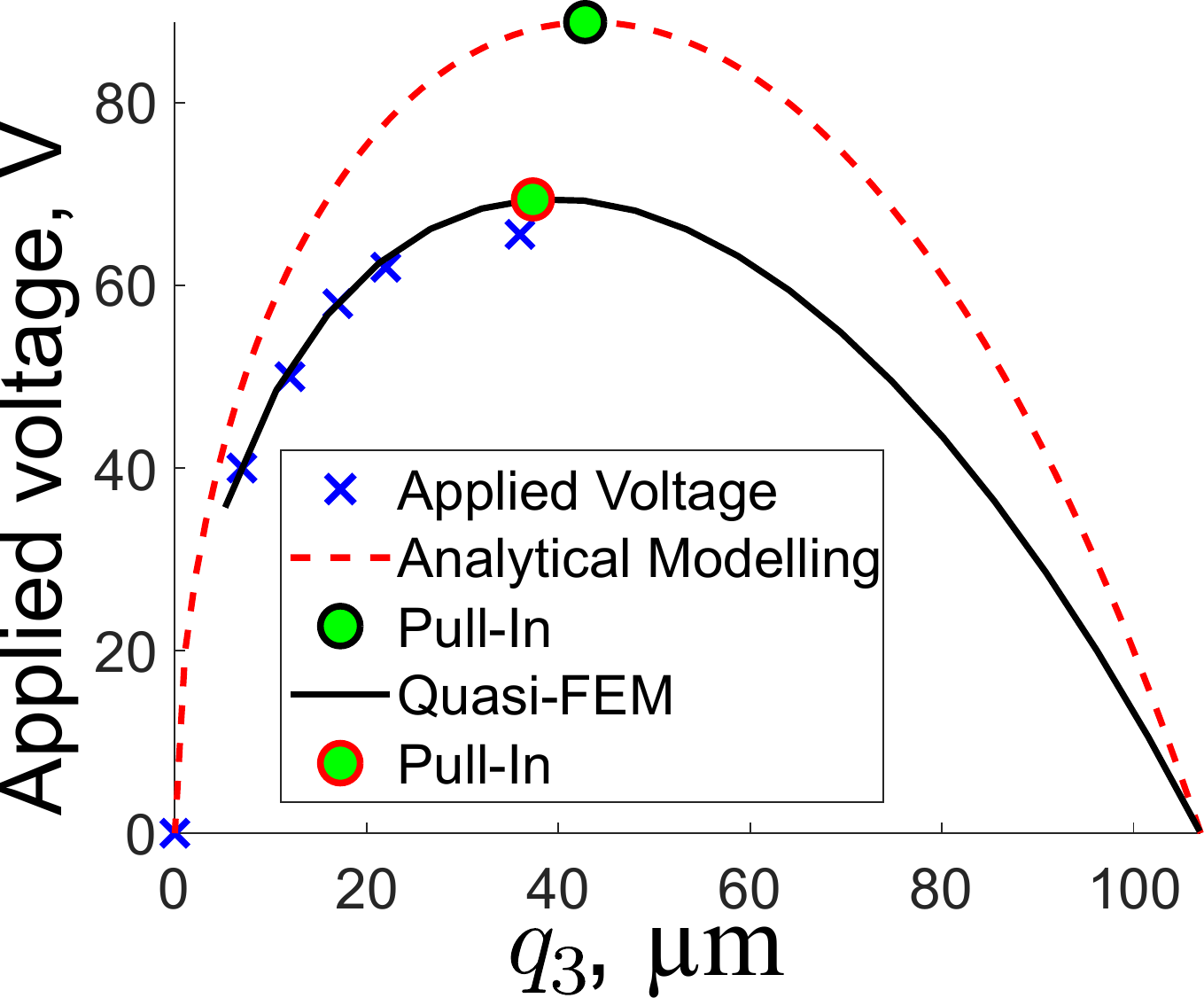}
  \caption{ Two equilibrium curves calculated by quasi-FEM Eq. (\ref{eq:model quasi})  and analytical model Eq. (\ref{eq:model I}) of applied voltage vs displacement of the disc having a  \SI{3.2}{\milli\meter} diameter and levitated at \SI{107}{\micro\metre} in comparing with experimental data.    }\label{fig:exper_disc 3_2mm_107_absolute}%\vspace*{-1.5em}
\end{figure}
\subsection{A light disc of a  \SI{2.4}{\milli\meter} diameter}
\label{sec:disc24}

Fig. \ref{fig:exper_disc 2_4mm}a) shows the result of the measurement of applied voltage to the electrodes 1 and 2 (see, Fig.\ref{fig:experement}b)) against the displacement of the disc having  a  \SI{2.4}{\milli\meter} diameter. The disc was levitated at a height of \SI{100}{\micro\meter} metering from the electrode plane and its displacement was measured by the laser sensor. The pull-in actuation occurred at a height of \SI{75}{\micro\meter} corresponding to the pull-in displacement of \SI{35}{\micro\meter}  upon applying \SI{38}{\volt} to the electrodes. Fig. \ref{fig:exper_disc 2_4mm}b) shows the comparison of equilibrium curves generated by quasi-FEM (\ref{eq:quasi pull-in dimensionless}) and analytical model (\ref{eq:analytic pull-in dimensionless})  with experimental measurement  in normalized voltage. For the simulation, the 3D geometrical scheme as shown in Fig. \ref{fig:3Dscheme} with same dimensions for coils are used. The disc is meshed by 3993 elements as shown in Fig. \ref{fig:mesh}.   The modelling is carried out with the following dimensionless parameters: $\xi=0.09$ and $\kappa=0.55$. Analysis of Fig. \ref{fig:exper_disc 2_4mm}b) depicts a good agreement between both developed models itself. In particular,  both  models predict the same value of the pull-in displacement. Only, we see a slight difference between the shapes of the curves  after the pull-in point. Also, both models in terms of normalized values are in good agreement with experiment.
However, the comparison in terms of dimension values shown in Fig. \ref{fig:exper_disc 2_4mm absolute} reveals that the analytical model gives a relative error, which is  around  \SI{16}{\%}  in estimation of the pull-in voltage.
\begin{figure}[!t]
  \centering
  \includegraphics[width=3.5in]{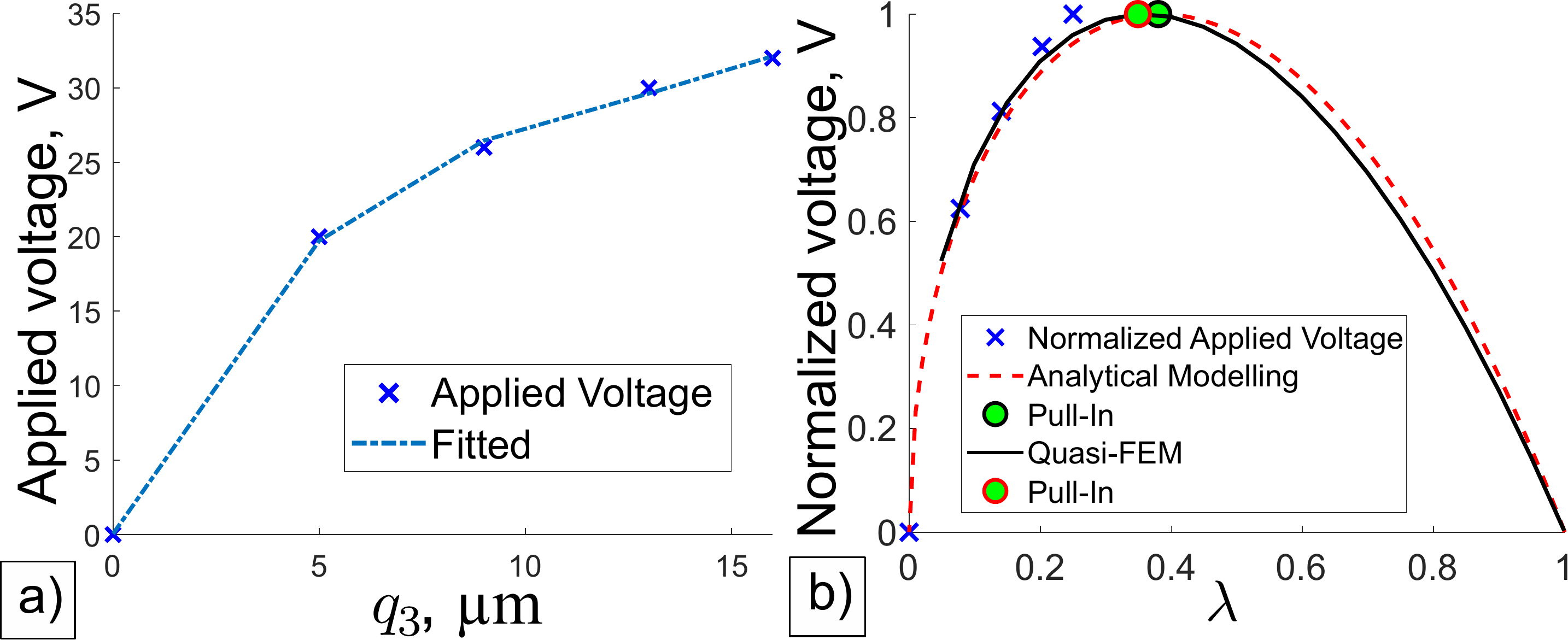}
  \caption{Pull-in actuation of disc having a  \SI{3.2}{\milli\meter} diameter and levitated at \SI{64}{\micro\metre}: a) experimental measurement of applied voltage vs linear displacement; b) normalized voltage  vs dimensionless displacement: measurement data together with equilibrium curves generated by quasi-FEM and analytical one.    }\label{fig:exper_disc 3_2mm_64}%\vspace*{-1.5em}
\end{figure}

\begin{figure}[!b]
  \centering
  \includegraphics[width=2.5in]{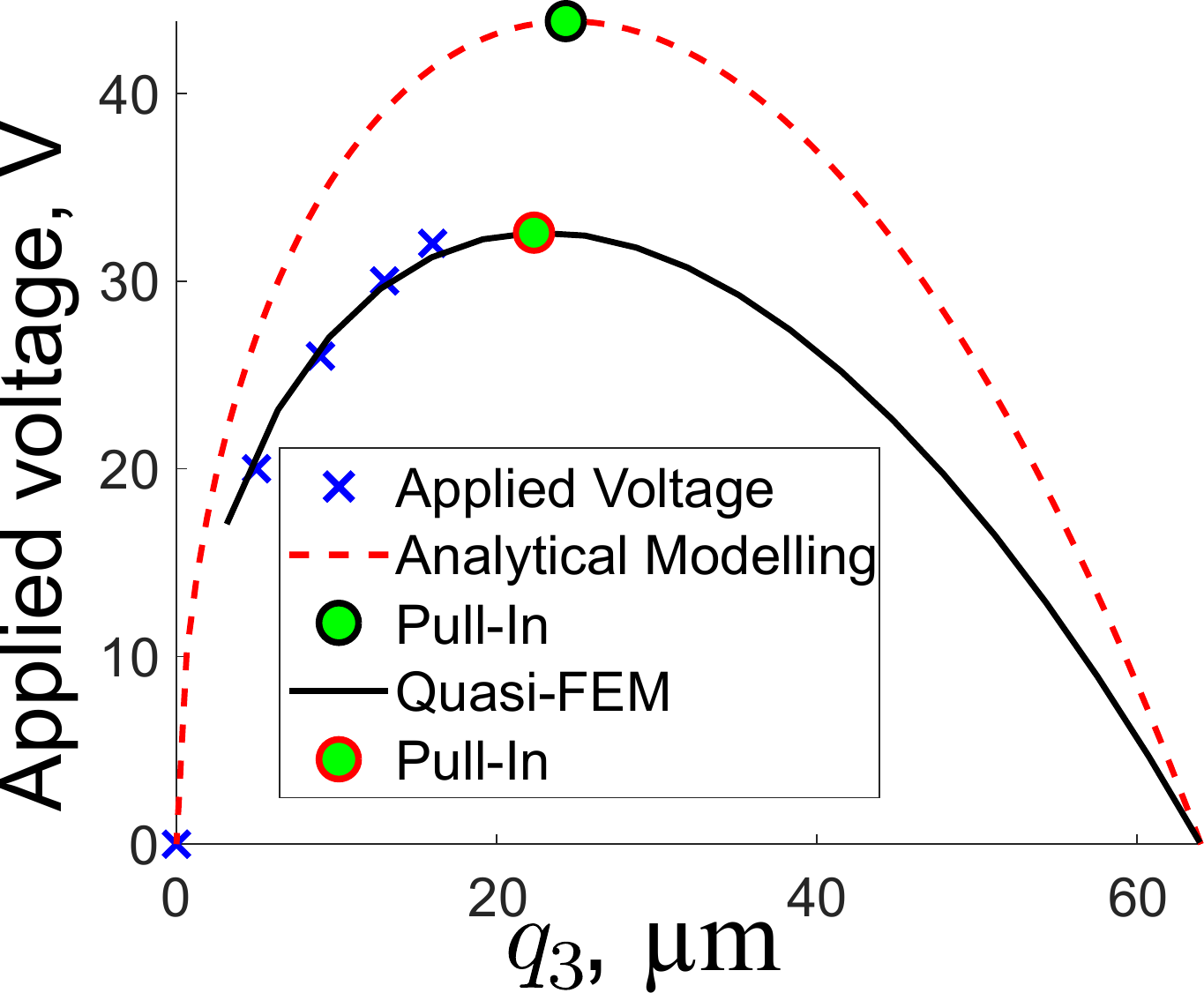}
  \caption{ Two equilibrium curves calculated by quasi-FEM Eq. (\ref{eq:model quasi})  and analytical model Eq. (\ref{eq:model I}) of applied voltage vs displacement of the disc having a  \SI{3.2}{\milli\meter} diameter and levitated at \SI{64}{\micro\metre} in comparing with experimental data.    }\label{fig:exper_disc 3_2mm_64_absolute}%\vspace*{-1.5em}
\end{figure}
\begin{table*}[t]
\caption{\label{tab}Results of measurements and modelling of
the static pull-in actuation.}
%\vspace*{-1.0em}
\begin{center}
\begin{tabular}{llcccc}
\hline
{Measured parameters}
            &Diameter of disc & \SI{2.4}{\milli\metre} &\SI{2.8}{\milli\metre}  &\multicolumn{2}{c}{\SI{3.2}{\milli\metre}} \\
     &Mass of disc & \SI{0.2}{\milli\gram} & \SI{0.3}{\milli\gram}  &  \multicolumn{2}{c}{\SI{0.7}{\milli\gram}} \\
%\hline
%\multirow{8}{*}{Measured parameters}
&Levitation height, $h_l$ &\SI{180}{\micro\metre} &\SI{200}{\micro\metre}&\SI{144}{\micro\metre}&\SI{187}{\micro\metre} \\
                    &Spacing, $h$ &\SI{100}{\micro\metre} & \SI{119}{\micro\metre}&\SI{64}{\micro\metre}&\SI{107}{\micro\metre} \\
                   % &  Diameter of levitation coil, $d_l$ & \SI{25}{\micro\metre} & \SI{25}{\micro\metre}&\\
\hline
\multirow{2}{*}{Calculated parameters}                     &$\xi=h_l/(2R_l)$&0.09&0.1&0.072&0.0935\\
     &$\kappa=h/h_l$&0.55&0.6&0.44&0.57\\
\hline
{\bf Measured pull-in   }                                    & {\bf Displacement} & \SI{35}{\micro\metre}  &\SI{43}{\micro\metre}&\SI{18}{\micro\metre}&\SI{36}{\micro\metre}\\
 {\bf  parameters }                 &  {\bf Voltage} & \SI{38}{\volt} & \SI{60.8}{\volt}&\SI{32}{\volt}&\SI{65}{\volt}\\

\hline
\hline
 {Pull-in parameters}
                            & {Displacement} & \SI{40}{\micro\metre} & \SI{49}{\micro\metre}&\SI{24}{\micro\metre} &\SI{42}{\micro\metre}\\
   modelled by Eq.(\ref{eq:model I})                         & {Voltage} & \SI{43}{\volt} & \SI{69}{\volt}&\SI{44}{\volt}&\SI{88}{\volt}\\
\hline
 {Pull-in parameters}
                          & {Displacement} & \SI{40}{\micro\metre} & \SI{48}{\micro\metre}& \SI{22}{\micro\metre}&\SI{37}{\micro\metre}\\
      simulated by Eq.(\ref{eq:model quasi})                     &  {Voltage} & \SI{37}{\volt} & \SI{60.76}{\volt}&\SI{33}{\volt}&\SI{69}{\volt}\\
\hline
\end{tabular}%\vspace*{-3.0em}
\end{center}
\end{table*}
\subsection{Disc of a  \SI{2.8}{\milli\meter} diameter}
Fig. \ref{fig:exper_disc 2_8mm}a) presents the result of the measurement of pull-in actuation of the disc having  a  \SI{2.8}{\milli\meter} diameter, which was obtained in our work \cite{Poletkin2015}. The disc was levitated at a height of \SI{120}{\micro\meter} measuring from the electrode plane.  The pull-in actuation occurred at a height of \SI{77}{\micro\meter} corresponding to the pull-in displacement of \SI{43}{\micro\meter}  upon applying \SI{60.8}{\volt} to the electrodes. The simulation is performed in a similar way as it was discussed in the previous section \ref{sec:disc24}. The difference is only in a size of diameter of the disc and a levitation height.  For modeling, the  following dimensionless parameters, namely, $\xi=0.1$ and $\kappa=0.6$ are used.
Comparison of both models in normalized values as shown in Fig. \ref{fig:exper_disc 2_8mm}b) depicts a  difference between the pull-in displacements predicted by   quasi-FEM (\ref{eq:quasi pull-in dimensionless}) and analytical model (\ref{eq:analytic pull-in dimensionless}). The analytical model (\ref{eq:analytic pull-in dimensionless})  gives a relative error, which is  around   \SI{2}{\%} in estimation of the pull-in displacement. Also, there is a slight difference between the shapes of the curves  after the pull-in point,  similar to the previous result.  Conducting the analysis in terms of dimension values as shown in Fig. \ref{fig:exper_disc 2_8mm absolute}, the relative error of around \SI{13}{\%} produced by the analytical model in the calculation of the pull-in voltage  can be recognized.

\subsection{Disc of a  \SI{3.2}{\milli\meter} diameter}

The result of the experimental investigation of the pull-in actuation in the presented prototype of HLMA with the disc of a  \SI{3.2}{\milli\meter} diameter was discussed and  reported in our work  \cite{PoletkinLuWallrabeEtAl2015}. Measuring from the electrode plane, the disc was levitated at a height of  \SI{107}{\micro\metre}. Fig. \ref{fig:exper_disc 3_2mm_107}a) shows the measurement of applied voltage to the electrodes against the disc linear displacement along the vertical axis.     The pull-in actuation occurred at a height of \SI{71}{\micro\meter} corresponding to the pull-in displacement of \SI{36}{\micro\meter}  upon applying \SI{65}{\volt} to the electrodes. The results  of simulation and modelling are shown in Fig. \ref{fig:exper_disc 3_2mm_107}b) in the normalized values. The analysis of Fig. \ref{fig:exper_disc 3_2mm_107}b) reveals a slight shift on the right of the equilibrium curve generated by the analytical model  (\ref{eq:analytic pull-in dimensionless}) with respect to the curve modeled by  the  quasi-FEM (\ref{eq:quasi pull-in dimensionless}).

Hence, this shift corresponds to the \SI{13.5}{\%} relative error in calculation of the pull-in displacement by the analytical model. The relative error produced by the analytical model in estimation of the pull-in voltage is calculated to be  \SI{30}{\%} as follows from  the analysis of the results of simulation and modelling in dimension values  shown in Fig. \ref{fig:exper_disc 3_2mm_107_absolute}.

In addition  we add new data of measurement of the pull-in actuation performed by the same disc, which was levitated  at a height of \SI{64}{\micro\metre}. The result of measurement of applied voltage against the linear displacement of the disc is shown in Fig. \ref{fig:exper_disc 3_2mm_64}. The pull-in actuation occurred at a height of  \SI{46}{\micro\metre} corresponding the pull-in displacement of   \SI{18}{\micro\metre}  upon applying the pull-in voltage equal to \SI{32}{\volt}. Similar to Fig. \ref{fig:exper_disc 3_2mm_107}b),  the equilibrium curve generated by the analytical model has a slight shift on the right relative to the curve generated by the quasi-FEM as shown in Fig. \ref{fig:exper_disc 3_2mm_64}b). This shift corresponds to the  \SI{9}{\%} relative error given by the analytical model in the estimation of the pull-in displacement.  Comparing in terms of dimension values as shown in Fig. \ref{fig:exper_disc 3_2mm_64_absolute}, the  relative error of around \SI{28}{\%} produced by the analytical model in the calculation of the pull-in voltage  is appeared.

The all results of measurements and modelling of the pull-parameters including related particularities of the experiments and values of dimensionless parameters for modelling are summed up in Table. \ref{tab}.

\section{Discussion and conclusion}
In this work,  the quasi-finite element method to model the static and dynamic behavior of  electromagnetic levitation micro-actuators based on the Lagrangian formalism has been developed.
The particularity of obtained quasi-finite element method    is the combination of finite element manner to calculate induced eddy current within a levitated micro-object and the set of  six differential equations describing the behavior of  mechanical part of electromagnetic levitation system. Using  a circular filament as a finite element for meshing the levitated micro-object, on the one hand, allows to covering any  shapes of the micro-object levitated by a system of wire-loops.  On the other hand, it allows to reducing the calculation of the mutual inductance between the  system of arbitrary shape wire loops and the levitated micro-object to a line integral as it was shown by Kalantarov and Zeitlin.
As a result, static and dynamic characteristics of an electromagnetic levitation system including the analysis of its stability can be  calculated and studied.

In particular, the developed method was applied to study static pull-actuation performed by the hybrid levitation micro-actuator. In general, the pull-in parameters of HLMA are a nonlinear function of its design due to the magnetic field generated by coil currents. However,  the developed model allows us to calculate accurately the pull-in parameters of HLMA for different diameters of disc levitated at different heights. The pull-in parameters are gathered under the following condition. Namely,  disc is meshed by 3993 elements and coils are represented by 32 circular filaments.  The time of calculation of one point was about \SI{120}{\second}. For building equilibrium curve, fifteen points were used.
The comparison of simulation results generated by the model (\ref{eq:quasi pull-in dimensionless}) based on quasi-finite element method with  the experiment  shows a good agreement  between the theory and measurement. This fact     confirms the efficiency of the developed method.

 At the same time, the analytical model (\ref{eq:analytic pull-in dimensionless}) based on the approximation of induced eddy current
within the disc by  one eddy current circuits was proposed as a result of the analysis of the distribution of the eddy current and the magnetic field generated by coil currents. Although, the effective use of the  analytical model requires the knowledge about the gradient of the magnetic field of a particular design under consideration.
However,   arising some particular cases, for instance,
when  electrodes, to generate electrostatic force acting on the disc, located very close to the disc bottom surface (dimensionless parameter, $\kappa$, is small, less than 0.1), the simple analytical equation (\ref{eq:simple pull-in dimensionless}) for estimation of the pull-in parameters with errors not exceeding  \SI{5}{\%} can be obtained.

% or


\appendices
\section{The derivative of dimensionless mutual inductance with respect to $\lambda$ }
\label{app}

Due to the particularity of the problem, namely, there is no  angular misalignment between a circular element and a coil. Hence, the
 original formula for calculation of the mutual inductance between two circular filaments based on the Kalantarov-Zeitlin approach \cite{Poletkin2019} in a dimensionless form can be rewritten  as follows:
 \begin{equation}\label{eq:NEW FORMULA}
  \overline{M}_{sj}=\frac{1}{\pi}\int_{0}^{2\pi}\frac{1+\overline{x}_1\cdot\cos\varphi+\overline{x}_2\cdot\sin\varphi}{\bar{\rho}^{1.5}}\frac{\Psi(k)}{k}d\varphi,
\end{equation}
where
\begin{equation}\label{eq:rho}
 \bar{\rho} =\sqrt{1+2(\overline{x}_1\cdot\cos\varphi+\overline{x}_2\cdot\sin\varphi)+\overline{x}_1^2+\overline{x}_2^2};
\end{equation}
\begin{equation}\label{eq:Psi}
  \Psi(k)=\left(1-\frac{k^2}{2}\right)K(k)-E(k),
\end{equation}
where $K(k)$ and $E(k)$ are the complete elliptic functions of the first and second kind, respectively;
\begin{equation}\label{eq:k}
   k^2=\frac{4\nu_j\bar{\rho}}{(\nu_j\bar{\rho}+1)^2+\nu_j^2\overline{x}_3^2},
\end{equation}
where $\nu_j=R_e/R_{cj}$, $R_{cj}$ is the radius of the $j$-coil filament, $\overline{x}_1$, $\overline{x}_2$ and $\overline{x}_3$ are the components of the radius vector $\boldsymbol{{r}}$ in base ${\boldsymbol{\underline{e}}^z}$ (see Eq. (\ref{eq:r(s,j)})).

The derivative of dimensionless mutual inductance with respect to $\overline{x}_3$  is
\begin{equation}\label{eq:mutual induction derivative of x3}
  \frac{\partial\overline{M}_{sj}}{\partial\overline{x}_3}=\frac{1}{\pi}\int_{0}^{2\pi}\frac{1+\overline{x}_1\cdot\cos\varphi+\overline{x}_2\cdot\sin\varphi}{\bar{\rho}^{1.5}}\Phi(k)d\varphi,
\end{equation}
where
\begin{equation}\label{eq:phi}
  \Phi(k)=\frac{d}{d\overline{x}_3}\frac{\Psi(k)}{k}=\frac{1}{k^2}\left(\frac{2-k^2}{2(1-k^2)}E(k)-K(k)\right)\frac{dk}{d\overline{x}_3},
\end{equation}
\begin{equation}\label{eq:k derivative of x3}
\frac{dk}{d\overline{x}_3}=-\frac{\nu_j^2\overline{x}_3\sqrt{4\nu_j\bar{\rho}}}{\left((1+\nu_j\bar{\rho})^2+\nu_j^2\overline{x}_3^2\right)^{3/2}}.
\end{equation}
Substituting  $\overline{x}_3=\lambda\kappa\chi$ into Eq. (\ref{eq:mutual induction derivative of x3}), the desired  equation for  the derivative of dimensionless mutual inductance with respect to $\lambda$ is derived.


\section*{Acknowledgment}

KP acknowledges with thanks the support from the priority programme
SPP 2206/1, the German Research Foundation (Deutsche Forschungsgemeinschaft, Grant KO 1883/37-1).
Also, KP deeply thanks Prof. U. Wallrabe for the continuous support of his research.

\ifCLASSOPTIONcaptionsoff
  \newpage
\fi

\bibliographystyle{IEEEtran}
\bibliography{References}

\end{document}